\documentclass{aastex62}

\usepackage{enumitem}
\usepackage{amsmath}
\newcommand\numberthis{\addtocounter{equation}{1}\tag{\theequation}}
\newcommand{\RNum}[1]{\uppercase\expandafter{\romannumeral #1\relax}}

\newcommand{\lya}{Ly$\alpha$}
\newcommand{\Rearth}{R$_{\oplus}$}
\newcommand{\Mearth}{M$_{\oplus}$}

\shorttitle{\sc K2-25b's Exosphere}
\shortauthors{Rockcliffe et al.}

\begin{document}

\title{\sc A Lyman-$\alpha$ transit left undetected: the environment and atmospheric behavior of K2-25b}

\author[0000-0003-1337-723X]{Keighley E. Rockcliffe}
\affiliation{Department of Physics and Astronomy, Dartmouth College, Hanover, NH 03755, USA}

\author[0000-0003-4150-841X]{Elisabeth R. Newton}
\affiliation{Department of Physics and Astronomy, Dartmouth College, Hanover, NH 03755, USA}

\author[0000-0002-1176-3391]{Allison Youngblood}
\affiliation{Laboratory for Atmospheric and Space Physics, University of Colorado Boulder, Boulder, CO 80303, USA}

\author[0000-0002-9148-034X]{Vincent Bourrier}
\affiliation{Observatoire Astronomique de l'Universit\'e de Gen\`eve, Chemin Pegasi 51b, CH-1290 Versoix, Switzerland}

\author[0000-0003-3654-1602]{Andrew W. Mann}
\affiliation{Department of Physics and Astronomy, University of North Carolina at Chapel Hill, Chapel Hill, NC 27599, USA}

\author[0000-0002-3321-4924]{Zachory Berta-Thompson}
\affiliation{Department of Astrophysical \& Planetary Sciences, University of Colorado Boulder, Boulder, CO 80303, USA}

\author[0000-0001-7077-3664]{Marcel A.~Ag\"ueros}
\affiliation{Department of Astronomy, Columbia University, New York, NY 10027, USA}

\author[0000-0002-8047-1982]{Alejandro N\'{u}\~{n}ez}
\affiliation{Department of Astronomy, Columbia University, New York, NY 10027, USA}

\author[0000-0002-9003-484X]{David Charbonneau}
\affiliation{Center for Astrophysics - Harvard-Smithsonian, Cambridge, MA 02139, USA}

\begin{abstract}

K2-25b is a Neptune-sized exoplanet ($3.45$\Rearth) that orbits its M4.5 host with a period of $3.48$ days. Due to its membership in the Hyades Cluster, the system has a known age ($727\pm 75$ Myr). K2-25b's youth and its similarities with Gl 436b suggested that K2-25b could be undergoing strong atmospheric escape. We observed two transits of K2-25b at Lyman-$\alpha$ using {\it HST}/STIS in order to search for escaping neutral hydrogen. We were unable to detect an exospheric signature, but placed an upper limit of $(R_{\text{p}}/R_{\star})|_{Ly\alpha} < 0.56$ at 95\% confidence by fitting the light curve of the Lyman-$\alpha$ red-wing, or $< 1.20$ in the blue-wing. We reconstructed the intrinsic Lyman-$\alpha$ profile of K2-25 to determine its Ly$\alpha$~flux, and analyzed {\it XMM-Newton} observations to determined its X-ray flux. Based on the total X-ray and extreme ultraviolet irradiation of the planet ($8763 \pm 1049$~erg~s$^{-1}$~cm$^{-2}$), we estimated the maximum energy-limited mass loss rate of K2-25b to be $10.6^{+15.2}_{-6.13} \times 10^{10}$ g s$^{-1}$ ($0.56$\Mearth~per 1 Gyr), five times larger than the similarly estimated mass loss rate of Gl 436b ($2.2\times 10^{10}$ g s$^{-1}$). The photoionization time is about 3 hours, significantly shorter than Gl 436b's 14 hours. A non-detection of a Lyman-$\alpha$ transit could suggest K2-25b is not significantly losing its atmosphere, or factors of the system are resulting in the mass loss being unobservable (e.g., atmosphere composition or the system's large high energy flux). Further observations could provide more stringent constraints.

\end{abstract}


\section{Introduction} \label{sec:intro}

Features in the exoplanet radius--period diagram are consequences of exoplanetary formation and evolutionary processes. In particular, the ``hot Neptune desert'' and the ``radius gap" motivate understanding atmospheric escape. The hot Neptune desert is the lack of short period planets (P$_{\text{p}}\lesssim3$ days) with radii between that of super-Earths and Jupiter \citep{2016NatCo...711201L,2016A&A...589A..75M,2014ApJ...783...54K,2013ApJ...763...12B,2011ApJ...727L..44S,2009MNRAS.396.1012D,2007A&A...461.1185L}. Atmospheric escape, along with orbital migration, is one of the dominant processes thought to shape the desert \citep{2018MNRAS.479.5012O,2016A&A...589A..75M}. The second feature, the radius gap, is the gap in the distribution of planetary radii around $1.5-2$\Rearth~\citep{2018AJ....156..264F,2017AJ....154..109F,2013ApJ...775..105O}. The radius gap can also be attributable to atmospheric escape \citep{2018MNRAS.476..759G,2014ApJ...795...65J,2013ApJ...776....2L,2013ApJ...775..105O}, though \cite{2021ApJ...908...32L} argue it can be primordial.

Atmospheric escape occurs on every planet with an atmosphere, but may have more or less influence on a its evolution depending on the planetary properties (e.g., bulk density, atmospheric composition, magnetic field) and environment (e.g., irradiation, stellar wind, impact erosion). The two atmospheric escape processes that have been mostly explored as shapers of the exoplanet population are thermal irradiation-driven (photoevaporation) and core-cooling-driven (core-powered mass loss) processes.

Photoevaporation occurs when a close-in planet with a volatile-dominated atmosphere receives a large flux of high energy radiation from its host. The radiation heats the planet’s upper atmosphere leading to the bulk motion of particles outward in a hydrodynamic outflow \cite[see][for a review]{2019AREPS..47...67O}. It follows that the timescale for this process is closely tied to the evolution of the star's high energy radiation. Although a timescale of 100 Myr is commonly accepted as it traces the period of highest stellar X-ray output \citep[e.g.,][]{2017ApJ...847...29O,2013ApJ...776....2L}, \cite{2021MNRAS.501L..28K} discuss the possibility of a longer Gyr timescale following the slower decline in extreme ultraviolet (EUV) output. Alternatively, \cite{2018MNRAS.476..759G} found that the luminosity from a cooling exoplanetary core (core-powered mass loss) can heat the atmosphere from below and cause hydrodynamic outflow. Exploring atmospheric escape within a range of host spectral types can distinguish between photoevaporation and core-powered mass loss \citep[e.g.,][]{2018MNRAS.476..759G}. These two processes have different dependencies on stellar and planetary properties, and different timescales for evolution. 

One of the most important tracers of atmospheric escape is the Lyman-$\alpha$ emission line (\lya; $1215.672$ \AA). A host star's \lya~radiation readily interacts with neutral hydrogen in the planetary atmosphere. While interstellar hydrogen usually completely absorbs \lya~at the emission line center, a planet transit can be observed at \lya~if the planetary hydrogen atoms are accelerated to high velocities such that they attenuate the wings of the line profile. The exact acceleration mechanisms remain an open question in the community. Atmospheric modelling work suggests that radiation pressure and stellar wind interactions change the spectral signature of the exosphere \citep{2020MNRAS.493.1292D,2019ApJ...873...89M,2018A&A...620A.147B,2016A&A...591A.121B,2015A&A...582A..65B,2018MNRAS.479.3115V,2013MNRAS.428.2565T,2010ApJ...709.1284B,2008Natur.451..970H}.

Neutral hydrogen exospheres in two transiting hot Neptunes, Gl 436b \citep[sometimes referred to as GJ 436b,][]{2019A&A...629A..47D,2017A&A...605L...7L,2015Natur.522..459E,2014ApJ...786..132K} and GJ 3470b \citep{2018A&A...620A.147B}, have been detected. These planets orbit their M dwarf hosts within $0.03$ AU and are therefore subject to high radiation levels. Observations with the {\it Hubble Space Telescope (HST)} Space Telescope Imaging Spectrograph (STIS) and the Cosmic Origins Spectrograph (COS) show deep and temporally asymmetric \lya~attenuation surrounding the white-light transit. These data suggest an extended comet-like hydrogen tail surrounding the planets.

\begin{deluxetable}{lcc}[t!]
\tablecaption{K2-25 system properties. \label{tab:prop}}
\tablecolumns{3}
\tablewidth{0pt}
\tablehead{
\colhead{Properties (Symbol)} &
\colhead{Value} &
\colhead{Units}
}
\startdata
Earth-system distance (d) & $45.014 \pm 0.165$ & pc \\
Age ($\tau$)$^{\text{a}}$ & $727 \pm 75$ & Myr \\
Right ascension ($\alpha$) & 04:13:05.62 & hh:mm:ss \\
Declination ($\delta$) & $+$15:14:51.9 & dd:mm:ss \\
Spectral type & M$4.5$ & \\
Bolometric luminosity (L$_{\text{bol}}$) & $8.16\pm 0.29 \times 10^{-3}$ & L$_{\odot}$ \\
Stellar mass (M$_{\star}$) & $0.2634 \pm0.0077$ & M$_{\odot}$ \\
Stellar radius (R$_{\star}$) & $0.2932 \pm0.0093$ & R$_{\odot}$ \\
Stellar rotation period (P$_{\star}$) & $1.88 \pm0.02$ & days \\
Barycentric radial velocity (v$_{\star}$) & $38.64 \pm0.15$ & km s$^{-1}$ \\
Epoch (t$_0$) & $2457062.57965 \pm 0.0002$ & BJD \\
Transit duration & $0.79^{+0.09}_{-0.17}$ & days \\
Planetary mass estimate$^{\text{b}}$ (M$_{\text{p}}$) & $7^{+10}_{-4}$ & \Mearth \\
Planetary mass measurement$^{\text{c}}$ & $24.5^{+5.7}_{-5.2}$ & \Mearth \\
Planetary radius (R$_{\text{p}}$) & $3.4492^{+0.1099}_{-0.1110}$ & \Rearth \\
Orbital period (P$_{\text{p}}$) & $3.48456322^{+9.7\times 10^{-7}}_{-9.5\times 10^{-7}}$ & days \\
Semi-major axis (a) & $0.0288 \pm0.0003$ & AU \\
\enddata
\tablecomments{Parameters are from \cite{2016ApJ...818...46M} and \cite{2020AJ....159...32T}.}
\tablenotetext{a}{Age estimate of the Hyades cluster as determined by \cite{2019ApJ...879..100D}.}
\tablenotetext{b}{Mass estimate from \cite{2020AJ....159...83K} using the mass-radius relation implemented in {\tt MRExo} \citep{2018ApJ...869....5N,2019ascl.soft12020K}.}
\tablenotetext{c}{Mass measurement from Habitable Zone Planet Finder radial velocity observations \citep{2020AJ....160..192S}}.
\end{deluxetable}
\begin{figure*}[t!]
    \centering
    \includegraphics[width=0.7\textwidth]{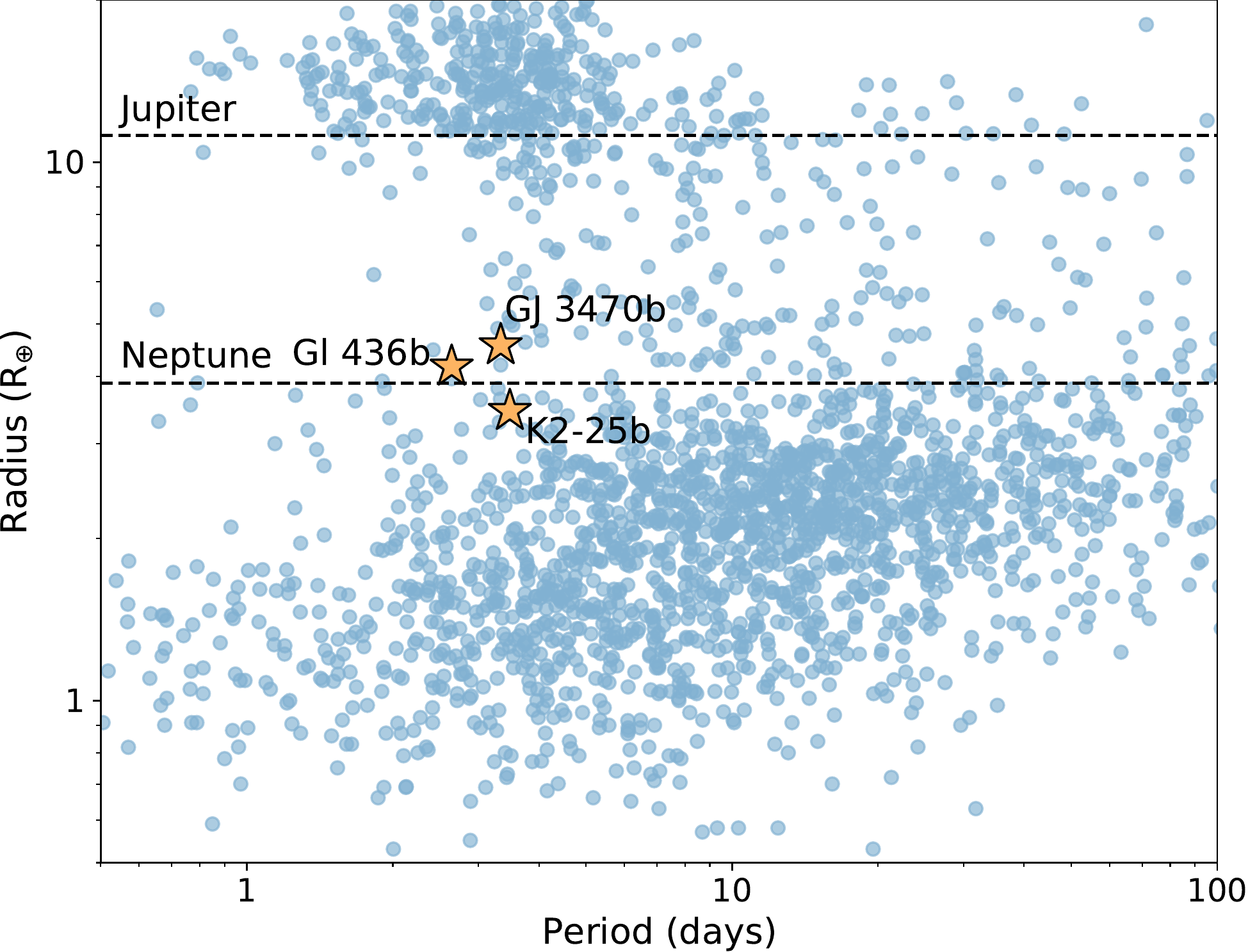}
    \caption{The distribution of radii (in Earth radii) as a function of orbital period (in days) for exoplanets within the NASA Exoplanet Archive as of March 2021.}
    \label{fig:valley}
\end{figure*}
K2-25b is a Neptune-sized short-period exoplanet within the $727 \pm 75$ Myr-old Hyades cluster \citep{2018ApJ...863...67G} that was discovered using photometry from the {\it K2} mission \citep{2016ApJ...818...46M}. K2-25b orbits around its young M4.5V host with a period of $3.48$ days. We review the physical properties of the K2-25 system in Table \ref{tab:prop}, and compare K2-25b to the Neptune-sized planets with detected neutral hydrogen exospheres in Figure~\ref{fig:valley}.

Determining how system properties like planetary atmospheric composition and youth impact mass loss is an important part of understanding atmospheric escape. Young exoplanets are particularly important for this study because they experience extreme stellar environments and directly probe the theorized $\sim100$ Myr timescale for photoevaporation. Constraining the mass loss rate and timescale for photoevaporation in young planets will test and improve exoplanet demographic studies. This motivates a detailed study of K2-25b and its radiation environment. \cite{2020MNRAS.498L.119G} previously analyzed infrared transmission spectra for K2-25b and did not detect escaping helium; here, we consider neutral hydrogen escape. We obtained {\it HST}/STIS \lya~observations of K2-25b (HST-GO-14615; PI: Newton), which we present in Section \ref{sec:obs}. Analysis of the \lya~light curves is presented in Section \ref{sec:lcurve}. We measure the \lya~and X-ray flux of K2-25, and estimate the energy-limited mass loss rate for K2-25b in Section \ref{sec:recon}. We conclude by discussing the implications of our results in Section \ref{sec:discuss}.


\section{Lyman-$\alpha$~line observations and analysis} \label{sec:obs}

We observed K2-25 with the $52$ x $0.1$\arcsec~aperture using {\it HST}'s STIS. We used the G140M grating with a spectral range of 1140-1741 \AA~and a resolving power of $\sim$10,000. We observed with the far-ultraviolet multi-anode microchannel array (FUV-MAMA) in TIME-TAG mode. Two {\it HST} visits occurred on 23 March 2017 (Visit 1) and 31 October 2017 (Visit 2), each corresponding to a transit of K2-25b. Eight science exposures were taken for each visit, each spanning the observable window of an {\it HST} orbit (about 2000 s). TIME-TAG mode observations individually stamp the arrival time for each photon detected by the instrument. These time-stamps allowed us to split the full exposures, prior to extraction and reduction, into three sub-exposures of $684$ s or $625$ s each.

\begin{figure}
    \centering
    \plotone{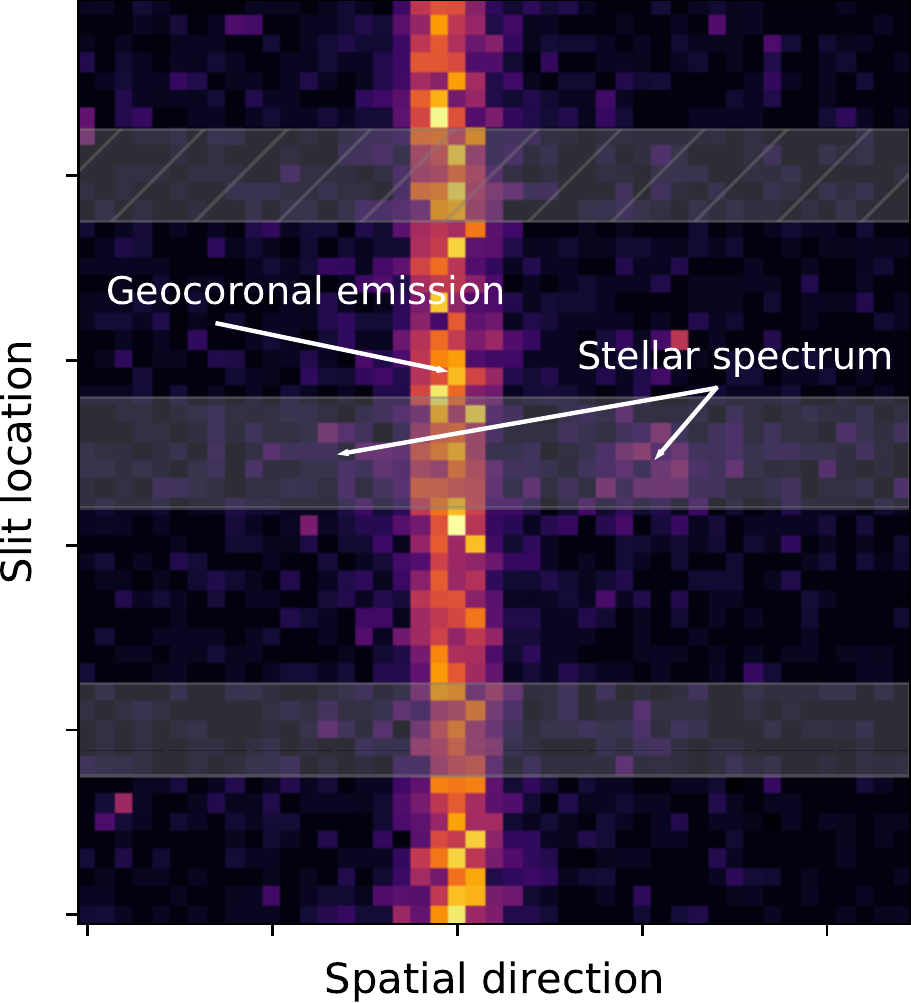}
    \caption{An example of the two-dimensional raw data taken by \textit{HST}/STIS during a full orbit within Visit 1. The spectral trace and the two background locations are highlighted in gray. In Visit 1, the geocoronal emission encroaches on the blue-wing of the \lya~line.}
    \label{fig:raw}
\end{figure}

We used the {\tt calstis} pipeline (v3.4) to reduce these data. We automatically located the extraction apertures by determining the centroid of the \lya~red wing. We used the {\tt calstis} pipeline to extract the background in two regions $15$ pixels above and below the extraction location. We used these extraction regions to fit the spatial direction with the default third-order polynomial, and this fit was used to remove the background from the target spectrum. We did not find the choice of polynomial order to be important. Cycle 27 calibration files were used to assign wavelengths to each pixel and then convert flux in counts to specific flux values (erg s$^{-1}$ cm$^{-2}$ \AA$^{-1}$).

Errors from {\tt calstis} are based on $\sqrt{\text{N}}$, which is inaccurate for small $N$ (the total counts). Equation \ref{eq:error} approximates the confidence limit for a Poisson distribution, corresponding to a Gaussian 1$\sigma$ limit \citep{1986ApJ...303..336G}. We used Equation \ref{eq:error} to recalculate errors for our spectra from the total counts (the {\tt GROSS} spectrum in the {\tt calstis} data products):
\begin{equation}
    \sigma \approx 1 - \sqrt{\text{N}+0.75} \label{eq:error}
\end{equation}
The error at each pixel was then converted to flux units using the flux conversion factors we inferred from the {\tt calstis} data products.

We looked for the {\it HST} ``breathing effect" (thermal changes in the telescope optics and focal plane) which can cause variability in flux throughput over the course of an {\it HST} orbit \citep{2013A&A...551A..63B,2012MNRAS.422.2477H,2008ApJ...686..658S,2001ApJ...552..699B,1997hstc.work...18B}. We created six light curves for the \lya\ line (three per visit) by grouping consecutive spectra. We folded the light curves on {\it HST}'s orbital period, and fit a sloped line to the flux as a function of orbital phase for each one. We also looked at the behavior of each exposure individually. In all cases, a sloped line fit and a horizontal line fit were similar as shown by their Bayesian Information Criteria ($|BIC_{slope} - BIC_{flat}| \sim 2$), so we did not include a systematics correction in our analysis.

\begin{figure*}[!t]
    \plotone{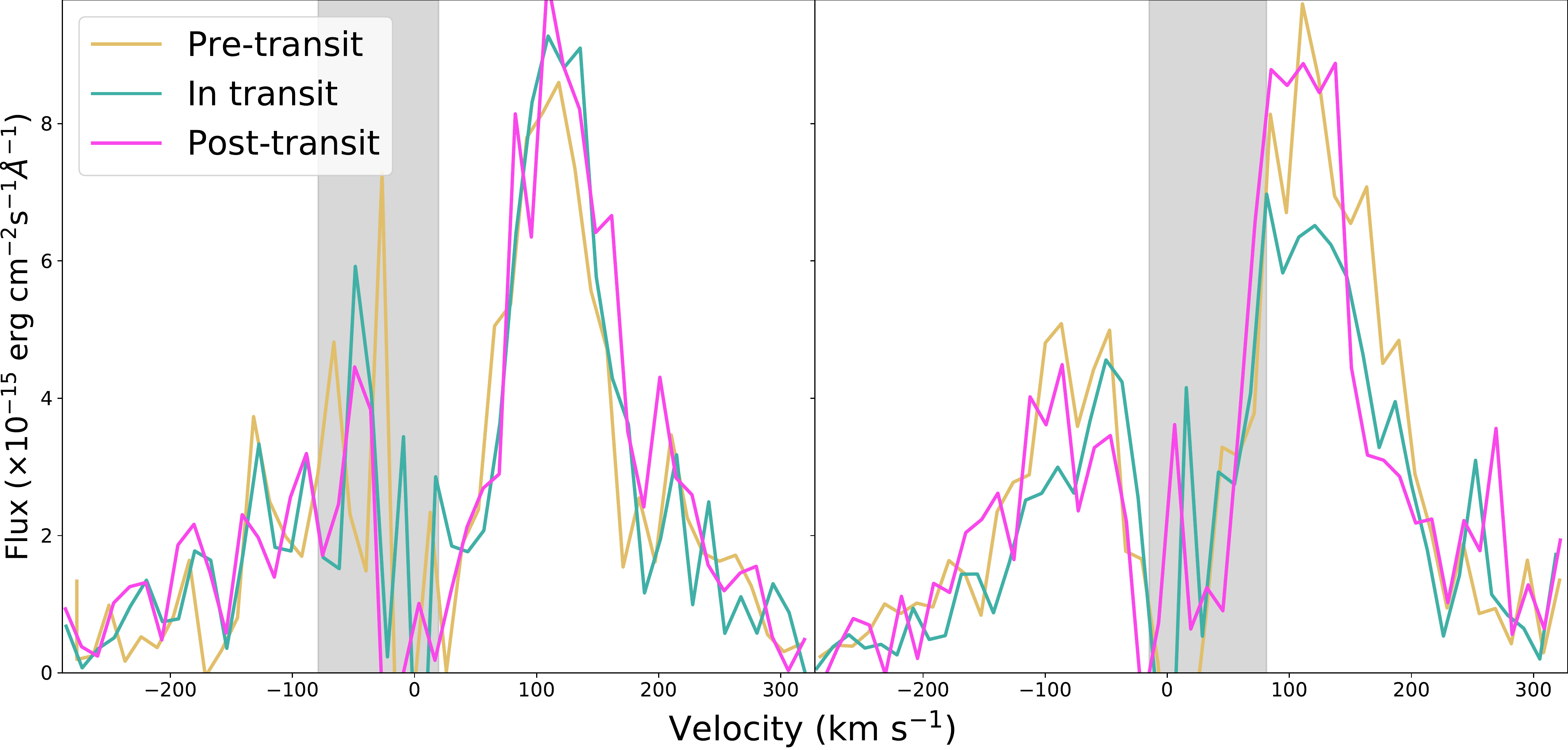}
    \caption{The spectra of K2-25 during Visits 1 (left) and 2 (right). The different colors indicate the average spectrum prior to transit (yellow), at times near the white-light transit (green), and after the transit (pink). The gray shaded region indicates the location of the geocoronal emission region that was removed from the spectra. Velocities are in the host star's reference frame.}
    \label{fig:specs}
\end{figure*}

The raw data are contaminated with geocoronal airglow from solar \lya~photons scattered off of the Earth's atmosphere (Figure \ref{fig:raw}). Though largely removed by background subtraction, contaminated regions should be treated with caution, as evidenced by the increased scatter in these regions as seen in Figure \ref{fig:specs} (gray regions). The contamination is of particular concern in Visit 1 because the geocoronal emission coincides with the blue-wing of K2-25's \lya~profile.

Given K2-25's systemic velocity, the \lya~line center is not observable at Earth due to the high neutral hydrogen column density, and for this reason we consider the \lya~wings. The individual spectra in the two panels of Figure \ref{fig:specs} temporally resolve each transit. If significant exospheric neutral hydrogen was present at high enough speeds, the wing of the \lya~profile would be attenuated during the exosphere transit. For comparison, one observation of the warm Neptune Gl 436b yielded a $56.2$\% transit in the blue-wing of \lya~starting 2 hours before and ending between 10 and 25 hours after the optical transit time of the planet \citep{2017A&A...605L...7L}. No deep transit is obvious for K2-25b during either Visit 1 or 2 (see Figure \ref{fig:specs}): all spectra taken during and outside of the planet transit are visually consistent.

\section{K2-25's Light Curve} \label{sec:lcurve}
\begin{figure}
    \centering
    \plotone{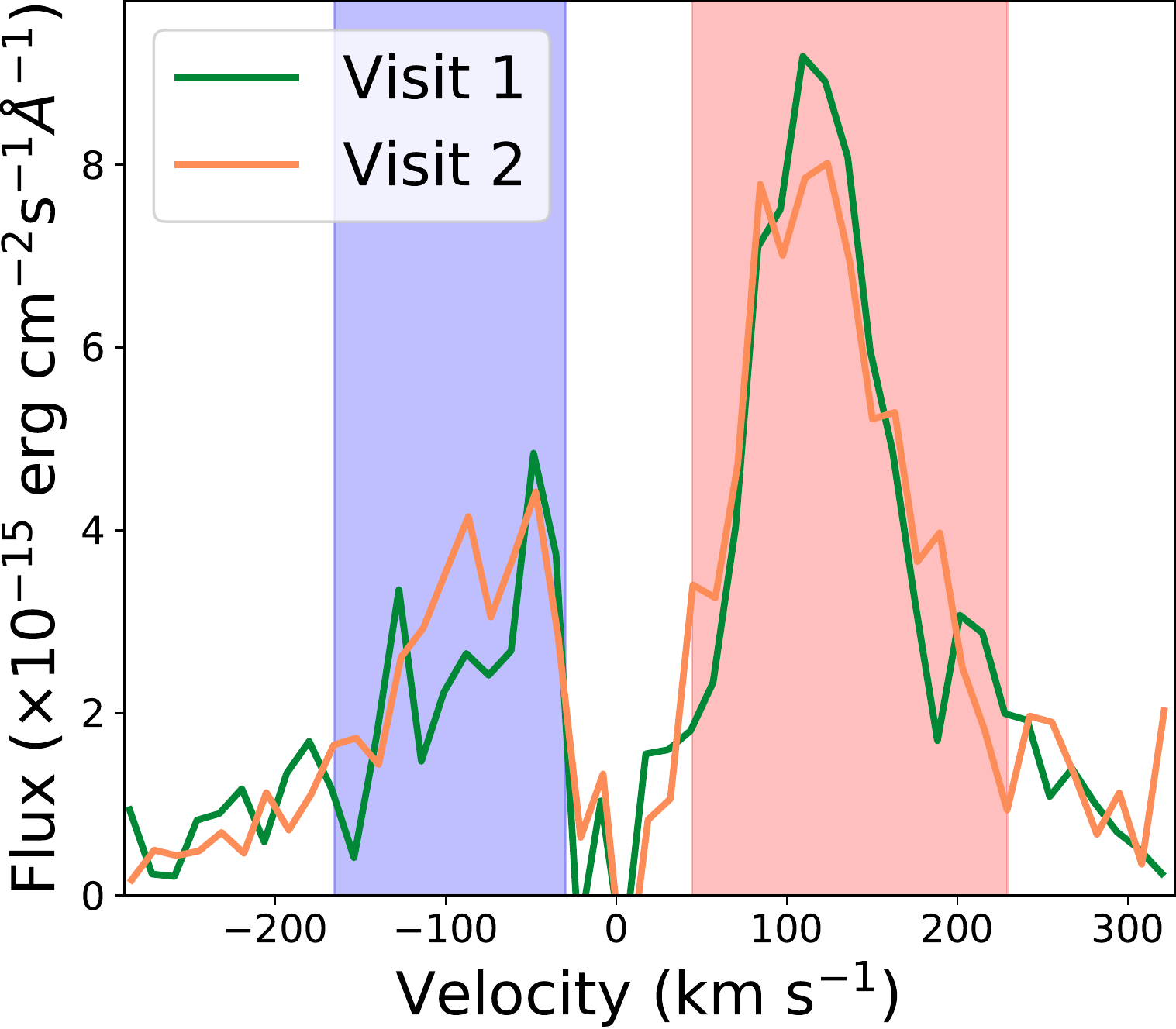}
    \caption{The \lya~average profiles for Visits 1 and 2. The integrated regions of the blue- and red-wing are shaded in blue and red, respectively. Velocities are in the host star's reference frame.}
    \label{fig:comp_specs}
\end{figure}

We created two \lya~light curves from our observations, one for the \lya~blue-wing and the other for the red-wing, in order to quantitatively investigate the presence of a transit. Our analysis uses the sub-exposures obtained from the TIME-TAG mode data. To obtain fluxes at each point in time, we summed over the stellar reference frame velocities $-165.2$ to $-29.5$ km s$^{-1}$ for the blue-wing, and $44.3$ to $229.3$ km s$^{-1}$ for the red-wing (Figure \ref{fig:comp_specs}).

We normalized all of the sub-exposure fluxes to the out-of-transit data using a linear fit to the full exposure data points that are beyond $\pm5$ hours of mid-transit. Both visits were combined into a single light curve as a function of planetary orbital phase (time from mid-transit). With sufficient signal and cadence, light curves such as these can constrain the radius and characterize the behavior of the potentially escaping neutral hydrogen.

We compared the \lya~red-wing light curve to K2-25's N~\RNum{5}~emission to look for evidence of non-transit related variability (e.g., flares). M dwarf flares enhance N \RNum{5}~flux more than \lya~and therefore should be more indicative of activity influencing our light curves \citep{2018ApJ...867...71L}. Two potential flares in N~\RNum{5}~were indicated in \cite{2020MNRAS.498L.119G}; improved error analysis does not support the presence of significant variability. Inclusion of the two points discarded in \cite{2020MNRAS.498L.119G} does not impact the results of that work. Since no flaring was evident in N~\RNum{5} or \lya~we used all data in our analysis.

We assumed the exosphere can be represented by a transiting opaque disk, and model the light curves with {\tt BATMAN} \citep{2015PASP..127.1161K}. The S/N of our data does not warrant considering more complicated transit shapes (e.g., asymmetry). While Gl 436b does produce a clear asymmetric transit shape, GJ 3470b's transit is relatively symmetric \citep[see Figures 2 and 5 in][]{2016A&A...591A.121B,2018A&A...620A.147B}. The full {\tt BATMAN} transit model consists of eight parameters: the mid-transit time ($t_0$), period ($P_{\text{p}}$), ratio of the planetary radius to stellar radius ($R_{\text{p}}/R_{\star}$), semi-major axis ($a$), inclination ($i$), eccentricity ($e$), argument of periastron ($\omega$), and the limb-darkening coefficients ($u$). We only fit for the transit depth, fixing all other parameters to the values listed in Table \ref{tab:batman}. The limb-darkening coefficients were assumed to be $0$, but are irrelevant given the sparse sampling of our data. 

There is a dip in the flux of both wings at $\sim1$ hour after the fixed transit time, lasting for one orbit. At this time, we do not explore asymmetric transits. Instead, we fit the same transit model, but vary the mid-transit time. The best fit results in a transit in both the blue and red-wings are centered at $\sim1$ hour which does not overlap with the transit ephemeris (the errors on the transit ephemeris are fractions of a minute, and \citet{2020AJ....159...83K} found no evidence of transit timing variations). As we are not aware of a model that would produce a signal like this, we assume this dip is not indicative of an exosphere and fix the mid-transit time to that expected from the white-light transit.

\begin{deluxetable}{cll}
\tablecaption{Light Curve Parameters.
\label{tab:batman}} 
\tablecolumns{3}
\tablewidth{0pt}
\tablehead{
\colhead{Parameter} &
\colhead{Value} &
\colhead{Units}
}
\startdata
$t_{0}^{\text{a}}$ & $0$ & hours \\
$P_{\text{p}}$ & $3.48456322$ & days \\
$a$ & $0.0288$ & AU \\
$i$ & $88.164$ & degrees \\
$e$ & $0.27$ & \\
$w$ & $98.0$ & degrees \\
\enddata
\tablenotetext{a}{The uncertainties on each transit epoch where we obtained observations are fractions of a minute, which allows us to fix this value.}
\end{deluxetable}

\begin{figure*}[t!]
    \centering
    \plotone{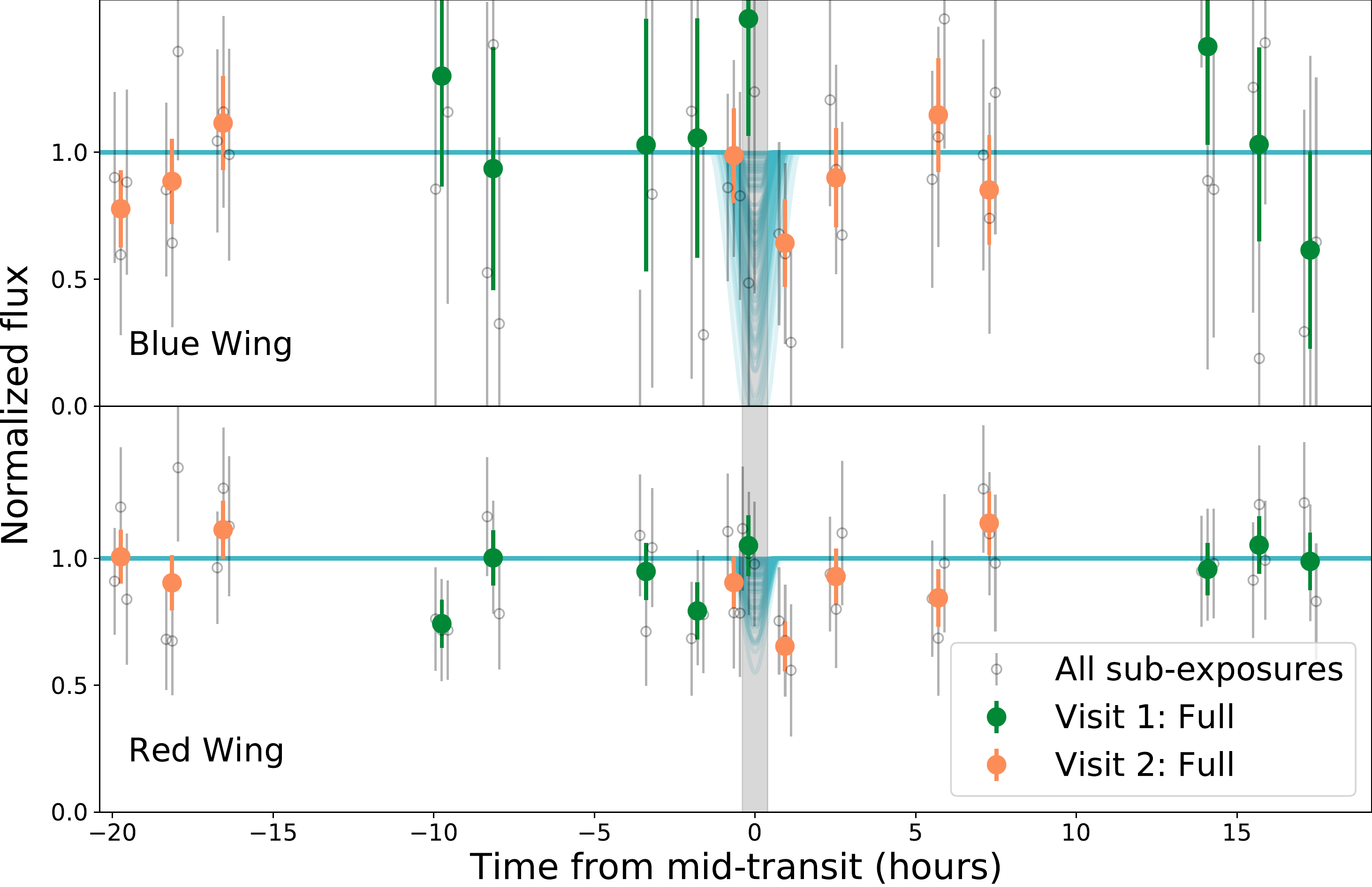}
    \caption{Light curves for \lya's blue-wing (top) and red-wing (bottom). Each panel has 16 filled circles, representing all of the full exposures from Visits 1 (green) and 2 (orange). The 48 empty circles are the sub-exposures. Samples from the posterior distributions are shown as light blue lines. The white-light transit time is shown by the gray shaded region.}
    \label{fig:lcurve}
\end{figure*}

The ratio of the planetary radius to the stellar radius was left to vary with a uniform prior and a lower limit of 0. We fit the model to the red and blue light curves using the Markov Chain Monte Carlo algorithm {\tt emcee} \citep{2013PASP..125..306F}. Figure \ref{fig:lcurve} shows the blue- and red-wing \lya~light curves and samples from the posterior distribution. The upper limits on $R_{\text{p}}/R_{\star}$ for the blue- and red-wing light curves are, respectively, $1.20$ and $0.56$ at $95$\% confidence.

\section{K2-25's High Energy Environment} \label{sec:recon}

Photoevaporation in the energy-limited regime is controlled by the entire X-ray to extreme ultraviolet (XUV) spectrum \citep[5-1170 \AA,][]{2012MNRAS.425.2931O,2009ApJ...693...23M,2004Icar..170..167Y,2003ApJ...598L.121L}. To investigate K2-25b's radiation environment, we determine the star's \lya, EUV, and X-ray flux. We then use these results to estimate the energy-limited mass-loss and photoionization rates for K2-25b.

\subsection{The intrinsic Lyman-$\alpha$ line}

We reconstructed the host's \lya~emission, modeling the intrinsic \lya~profile and the attenuation from the interstellar medium (ISM). We based our reconstruction on the combined spectrum from all {\it HST} orbits because there was no planetary signal in the data. To produce the combined spectrum, we performed a weighted average of all 16 spectra. We weighted the spectra by the associated error, using an average ``error spectrum'' for the two visits, which was applied to all eight spectra from that visit. We used this method to avoid biasing fluxes to lower values. While the error bars for the spectra from within each visit are consistent with each other, the changing location of the geocoronal emission means the errors changes between the two visits (see Figure \ref{fig:specs}). We compute the average error spectrum by averaging the errors at each pixel for the eight spectra from a single visit.

We used the model developed in \cite{2016ApJ...824..101Y} that describes the transmission profile of a local low-mass star with nine parameters. The model and fitting algorithm are available in the Python package {\tt lyapy}\footnote{https://github.com/allisony/lyapy}. 

The emission profile was composed of a narrow and broad Gaussian, each characterized by an amplitude (A$_{\text{n}}$ and A$_{\text{b}}$), full-width half-maximum (FW$_{\text{n}}$ and FW$_{\text{b}}$), and heliocentric velocity centroid (v$_{\text{n}}$ and v$_{\text{b}}$). The interstellar medium attenuation was modeled with a Voigt profile approximation \citep{1948ApJ...108..112H} described by a Doppler broadening parameter ($b$), line of sight \ion{H}{1} column density (N$_{\text{H I}}$), and velocity centroid (v$_{\text{H I}}$). The interstellar medium's \ion{D}{1} content was characterized by the D/H ratio, which was left to vary. The velocity centroid for deuterium absorption was assumed to be the same as neutral hydrogen.

This model assumes that all of the interstellar neutral hydrogen exists in one cloud. While the LISM Dynamical Model \citep{2008ApJ...673..283R} outputs three clouds in the direction of K2-25, the low S/N of our spectra are not sufficient to constrain multiple ISM components \citep{2016ApJ...824..101Y}.

\begin{deluxetable*}{llll}
\tablecaption{The best-fit parameter values with 1$\sigma$ uncertainties for the intrinsic \lya~profile. \label{tab:bfvals}} 
\tablecolumns{2}
\tablewidth{0pt}
\tablehead{
\colhead{Parameter} &
\colhead{Value} &
\colhead{Units} &
\colhead{Description}
}
\startdata
v$_{\text{n}}$ & $38.19^{+2.70}_{-2.17}$ & km s$^{-1}$ & narrow component velocity centroid \\
log$_{10}$A$_{\text{n}}$ & $-13.75^{+0.08}_{-0.07}$ & erg s$^{-1}$ cm$^{-2}$ \AA$^{-1}$ & narrow component amplitude \\
FW$_{\text{n}}$ & $163.40^{+15.64}_{-16.27}$ & km s$^{-1}$ & narrow component FWHM \\
v$_{\text{b}}$ & $38.56^{+2.87}_{-2.45}$ & km s$^{-1}$ & broad component velocity centroid \\
log$_{10}$A$_{\text{b}}$ & $-14.72^{+0.21}_{-0.26}$ & erg s$^{-1}$ cm$^{-2}$ \AA$^{-1}$ & broad component amplitude \\
FW$_{\text{b}}$ & $447.17^{+178.28}_{-78.07}$ & km s$^{-1}$  & broad component FWHM \\
log$_{10}$N$_{\text{H I}}$ & $18.14^{+0.08}_{-0.08}$ & & ISM H I column density \\
b & $8.62^{+3.22}_{-4.01}$ & km s$^{-1}$ & ISM Doppler broadening parameter \\
v$_{\text{H I}}$ & $12.91^{+2.83}_{-2.80}$ & km s$^{-1}$ & ISM velocity centroid \\
D/H & $6.95^{+6.40}_{-4.37}\times 10^{-6}$ & & ISM deuterium-to-hydrogen ratio \\
\enddata
\end{deluxetable*}

We fit the stellar emission and ISM attenuation to the observed \lya~spectrum ($-350<v<390$ km s$^{-1}$). {\tt lyapy} uses the MCMC sampler \citep[{\tt emcee};][]{2013PASP..125..306F} to explore the parameter space. The chains were initialized by randomly sampling a normal distribution. The Doppler broadening parameter had a logarithmic prior. Uniform priors were assumed for all other varied parameters, including the column density (N$_{\text{H I}}$). The posterior distributions were sampled 100,000 times with a burn-in of 20,000 for 30 walkers.

We took the median of each one-dimensional posterior distribution (marginalized distribution) as the best-fit value and defined the uncertainty by the 16th and 84th percentiles of the distribution. The best-fit values and their uncertainties are listed in Table \ref{tab:bfvals}. The best-fit column density, log$_{10}$N$_{\text{H I}} = 18.14^{+0.08}_{-0.08}$, is consistent with the column density obtained by the Colorado Model of the Local Interstellar Cloud \citep{2000ApJ...534..825R} for the line of sight to K2-25, log$_{10}$N$_{\text{H I}} = 18.17$. There are two other clouds (Hyades, Aur) that are predicted to lie along K2-25's sight line \citep{2008ApJ...673..283R} which could account for the differences between our ISM results and the literature. The best-fit deuterium-to-hydrogen ratio (D/H$ = 6.95^{+6.40}_{-4.37}\times 10^{-6}$) has large enough uncertainties such that it is consistent with the accepted literature value of $1.5\times 10^{-5}$ found by \cite{2006ApJ...647.1106L} (see also \citealt{2004ApJ...609..838W,2003ApJ...599..297H}). The large uncertainties in the flux within the profile region where deuterium absorption occurs ($\sim -100$ to $-40$ km s$^{-1}$) result from the overlap with the geocoronal emission region in Visit 1 (see Figure~\ref{fig:specs}).

\begin{figure*}[t!]
    \centering
    \includegraphics[width=0.7\textwidth]{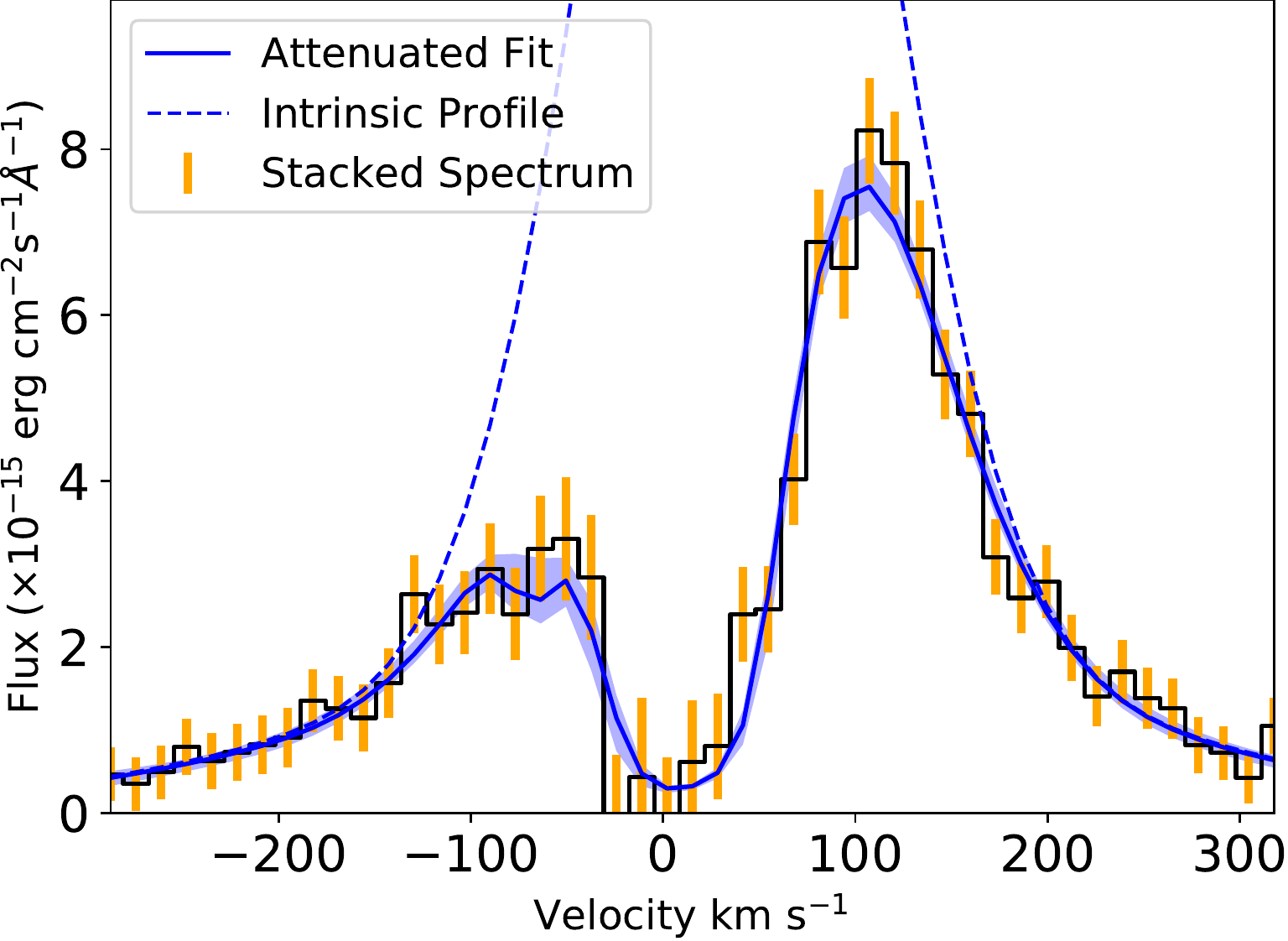}
    \caption{Our reconstructed \lya~profile for K2-25 (dashed blue). The spectrum created from the sixteen stacked exposures is included in black, with errors in orange. The best-fit profile - a combination of the \lya~double component Gaussian profile and the ISM's attenuating Voigt profile - is depicted as the solid blue line with 1$\sigma$ errors shown in shaded blue.}
    \label{fig:prof}
\end{figure*}

K2-25's reconstructed \lya~profile is shown in Figure \ref{fig:prof}. K2-25's \lya~flux at 1 AU from the star is $1.38^{+0.18}_{-0.13}$ erg s$^{-1}$ cm$^{-2}$, where the uncertainties are $1\sigma$ error bars.

\subsection{The EUV spectrum}

We estimated K2-25's EUV spectrum ($100-1170$ \AA) using empirical relations from \cite{2014ApJ...780...61L}. Using updated solar upper atmospheric models from \cite{2014SoPh..289..515F}, \cite{2014ApJ...780...61L} determined that the relationship between EUV and \lya~flux is roughly constant with \lya~flux. They obtained EUV to \lya~flux ratios in nine wavebands from $100-1170$ \AA~(indicated in Table \ref{tab:XEUV}). These relations are accurate within $20$\%, as inferred from the spectra of F5 V - M5 V stars observed with the {\it Extreme UltraViolet Explorer} and {\it Far Ultraviolet Spectroscopic Explorer} \citep{2014ApJ...780...61L}. Each waveband was summed and scaled to calculate the total EUV flux of K2-25 at 1 AU, $1.40 \pm 0.14$ erg s$^{-1}$ cm$^{-2}$.

There may be additional, unaccounted for, systematic errors in the EUV spectrum for K2-25 because stars of its mass and age have not been thoroughly investigated by the models and observations. For example, \cite{2019ApJ...886...77P} showed that the EUV fluxes for Gl 436 from synthetic spectra and from the relations within \cite{2014ApJ...780...61L} can differ by a factor of a few.

\subsection{The X-ray spectrum}

K2-25 was observed with {\it XMM-Newton} for a broad investigation into stellar rotation and activity (OBSID 0782061001; PI: Ag\"ueros). The spectrum covers $5-100$~\AA~($0.12-2.48$ keV). We fit the spectrum output from the {\it XMM-Newton}/MOS1 instrument with a one temperature VAPEC model, which characterizes the emission associated with collisionally-ionized gas (e.g., stellar coronae). The fit was performed with the HEASARC software {\tt xspec}\footnote{https://heasarc.gsfc.nasa.gov/xanadu/xspec/}, an X-ray spectral fitting package. The fitted spectrum was integrated to get a total soft X-ray flux with 1$\sigma$ uncertainty of $5.88^{+1.16}_{-0.56}$~erg~s$^{-1}$~cm$^{-2}$ at 1~AU from the star. 
The ratio of K2-25's X-ray luminosity to its bolometric luminosity is $\log_{10}{L_\text{X}/L_{\text{bol}}}=-3.28$.

K2-25's Rossby number, defined as the ratio of the rotation period to its convective turnover time, is 0.02 according to the mass-convective turnover time relation from \citet{2018MNRAS.479.2351W}. Its Rossby number places K2-25 in the saturated part of the rotation-coronal activity relation, where the X-ray emission appears to decouple from a star's rotation rate and is roughly constant. For 20 Hyades members with $M_*<$1.2 M$_{\odot}$ and X-ray detections that fall in the saturated regime, $\log_{10}{L_\text{X}/L_{\text{bol}}}=-3.17^{+0.11}_{-0.16}$ (Núñez et al., in prep.), with uncertainties corresponding to the 16th and 84th percentiles. K2-25's X-ray flux is fully consistent with that of other young, rapidly rotating stars.

Gl 436 and GJ 3470 are field-age stars and have longer rotation periods of 44.09 days and 21.54 days, respectively \citep{2018Natur.553..477B,2019AJ....157...97K}. \cite{2018ApJ...863..166V} were able to constrain Gl~436's age to $8.9^{+2.3}_{-2.1}$ Gyr. Gl 436 and GJ 3470 have estimated Rossby numbers of 0.78 and 0.47, using the mass-convective turnover time relation from \citet{2018MNRAS.479.2351W} and masses from \citet{2019AJ....158..138S}. These Rossby numbers are well within the unsaturated regime. K2-25 has an X-ray luminosity that is ten times larger than GJ 3470 and one hundred times larger than Gl 436, which is compatible with K2-25's youth and corresponding rapid rotation.

\subsection{The XUV spectrum}

The X-ray to EUV fluxes and errors are reported in Table \ref{tab:XEUV}. The sum of the X-ray and EUV (XUV) irradiation that K2-25b experiences is $8763 \pm 1049$~erg~s$^{-1}$~cm$^{-2}$, where the error was calculated using the average X-ray flux uncertainty and $20\%$ uncertainty in each EUV band. This is roughly $70\%$ of the XUV flux estimated by \cite{2020MNRAS.498L.119G}, who based their estimate on a scaled composite spectrum of GJ 674 and rotation-activity relations, and is consistent with their uncertainty (see their Table 1). K2-25b's XUV irradiation is roughly seven times that of Gl 436b and twice that of GJ 3470b \citep{2018A&A...620A.147B,2016A&A...591A.121B}.
\begin{deluxetable}{ccc}
\tablecaption{K2-25's X-ray and EUV emission.\label{tab:XEUV}}
\tablecolumns{2}
\tablewidth{0pt}
\tablehead{
\colhead{Waveband (\AA)} & \colhead{Flux at 1 AU} & \colhead{Uncertainty} \\
\colhead{} & \colhead{(erg s$^{-1}$ cm$^{-2}$)} & \colhead{(erg s$^{-1}$ cm$^{-2}$)}
}
\startdata
$5-100$ & $5.883$ & $+1.159$/$-0.563$ \\
$100-200$ & $0.444$ & $\pm0.089$ \\
$200-300$ & $0.389$ & $\pm0.079$ \\
$300-400$ & $0.344$ & $\pm0.069$ \\
$400-500$ & $0.008$ & $\pm0.002$ \\
$500-600$ & $0.013$ & $\pm0.003$ \\
$600-700$ & $0.018$ & $\pm0.004$ \\
$700-800$ & $0.021$ & $\pm0.004$ \\
$800-912$ & $0.027$ & $\pm0.005$ \\
$912-1170$ & $0.139$ & $\pm0.028$ \\
\lya & $1.375$ & $+0.182$/$-0.126$ \\
\enddata
\end{deluxetable}
\\
\subsection{Mass loss rate}

The mass measurement for K2-25b from \cite{2020AJ....160..192S} indicates its atmosphere is currently dominated by hydrogen, and is therefore still susceptible to hydrodynamic escape. We used our XUV flux to estimate the total mass loss rate of K2-25b, adopting the energy-limited methodology as reviewed in \cite{2019AREPS..47...67O}. The energy-limited approach was originally derived by \cite{1981Icar...48..150W} using outflow calculations for a young Earth and Venus. The equation is: 
\begin{equation} \label{eq:loss}
    \dot{M} = \eta \frac{\pi R^3_{\text{p}}F_{\text{XUV}}}{GM_{\text{p}}K_{\text{eff}}}
\end{equation}
\begin{equation} \label{eq:Keff}
    K_{\text{eff}} = \frac{(a/R_{\text{p}} -1)^2 (2a/R_{\text{p}} +1)}{2(a/R_{\text{p}})^3}
\end{equation}
where $R_{\text{p}}$ is the planetary radius, $F_{\text{XUV}}$ is the stellar XUV flux at the planet's location, and $M_{\text{p}}$ is the planetary mass. This formalism assumes that K2-25b is absorbing all of the XUV flux at its optical-wavelength radius, where the energy input balances the energy necessary for escaping the planet's potential.

Two correction factors remain in Equation~\ref{eq:loss}. \cite{2007A&A...472..329E} added the $K_{\text{eff}}$ factor (Equation \ref{eq:Keff}) to correct for the potential difference between the planet's radius and its Roche lobe, which needs to be overcome by the energetic atmosphere in order to escape. The $\eta$ factor characterizes the efficiency at which the XUV radiation heats the planet's atmosphere. The exoplanet population could exhibit a wide range of heating efficiencies (0.15-0.6) \citep{2014ApJ...795..132S}. $\eta$ has been shown to depend on the incident flux level -- higher flux levels cause radiative cooling to become a limiting factor \citep{2009ApJ...693...23M} -- and atmospheric properties. We report our mass loss rate in terms of $\eta$, (i.e, without assuming a heating efficiency), which enables comparison to other systems.

We calculated the total mass loss rate to be $10.6^{+15.2}_{-6.13}\eta \times 10^{10}$ g s$^{-1}$ assuming the planetary parameters in Table \ref{tab:prop} and the estimated planetary mass derived from mass-radius relations. This is about 50\% of the estimate from \cite{2020MNRAS.498L.119G}, who use a different method for calculating total XUV irradiation, but agrees within our uncertainty.

\cite{2020AJ....160..192S} used radial velocity observations from the Habitable-zone Planet Finder to measure K2-25b's mass, a challenging prospect given that the $1.8$ day stellar rotation period is close to half the planetary orbital period. These authors determine a mass of $24.5^{+5.7}_{-5.2}$\Mearth, higher than our estimate of $7$\Mearth~and the estimate of $9.7$\Mearth~from \cite{2020MNRAS.498L.119G}. \footnote{They also combined {\it K2} and ground-based photometry observations to obtain a planetary radius of $3.44 \pm 0.12$\Rearth, in agreement with the value from \cite{2020AJ....159...32T} and quoted in Table 1.} Using their planet mass and radius gives a mass loss rate of $3.00^{+0.77}_{-0.71}\eta \times 10^{10}$ g s$^{-1}$. This is about 30\% of the estimate using the parameters from Table \ref{tab:prop} which is within our uncertainty and does not substantively alter our conclusions (see Section \ref{sec:discuss}).

\subsection{Photoionization rate}

The XUV irradiation of K2-25b photoionizes its exosphere. The average lifetime of an escaping neutral hydrogen atom contributes to whether or not an absorption signature should be observable in \lya. Equation \ref{eq:photrate} gives the photoionization rate per neutral hydrogen atom given an incident flux \citep[see e.g.][]{2016A&A...591A.121B}:
\begin{equation} \label{eq:photrate}
    \Gamma_{\text{ion}} = \int^{911.8\text{ \AA}}_{0\text{ \AA}} \frac{F_{\text{XUV}}(\lambda)\sigma_{\text{ion}}(\lambda)}{hc}\lambda d\lambda
\end{equation}
where:
\begin{align*}
    \sigma_{\text{ion}} = &6.538\times 10^{-32} \left(\frac{29.62}{\sqrt{\lambda}}+1\right)^{-2.963} \\
    & \times (\lambda-28846.9)^2 \lambda^{2.0185} \numberthis \label{eq:sigion}
\end{align*}
The photoionization rate, $\Gamma_{\text{ion}}$, is given in s$^{-1}$. $F_{\text{XUV}}$ is the stellar XUV flux at the location of the planet in erg s$^{-1}$ cm$^{-2}$ \AA$^{-1}$, and $\sigma_{\text{ion}}$ is the photoionization cross section in cm$^2$ given in equation \ref{eq:sigion} \citep{2013ccfu.book...67B,1996ApJ...465..487V}. The integral is over the H-ionizing portion of the electromagnetic spectrum ($\lambda \leq 911.8$ \AA).

We calculated the photoionization rate by summing over the XUV flux values listed in Table \ref{tab:XEUV} multiplied by $\sigma_{\text{ion}}(\lambda)\lambda$ evaluated at the average wavelength of each wavelength bin. This results in a photoionization rate of $9.73\times 10^{-5}$ s$^{-1}$ at K2-25b's semi-major axis, which corresponds to a neutral hydrogen atom's lifetime of $2.86$ hours. GJ 3470b was similarly found to have a short neutral hydrogen lifetime of $\sim 0.9$ hours \citep{2018A&A...620A.147B}. In contrast, Gl 436b has a long neutral hydrogen lifetime of about $14$ hours which allows the leading and trailing regions of its exosphere to evade rapid ionization \citep{2016A&A...591A.121B}. Short neutral hydrogen lifetimes (large photoionization rates) result in smaller neutral exospheres and correspondingly shallower and shorter \lya~transits \citep[as can be seen by comparing the Gl 436b and GJ 3470b transit signals;][]{2016A&A...591A.121B,2018A&A...620A.147B}.


\section{Discussion} \label{sec:discuss}

Observations of young planets inform our knowledge of exoplanetary evolution, and the impact of host star age and activity on planetary properties.

\subsection{Summary}

K2-25b is similar in size and proximity to its host star ($3.45$\Rearth, $0.0288$ AU) to Gl 436b ($4.26$\Rearth, $0.0287$ AU) and GJ 3470b ($4.57$\Rearth, $0.0355$ AU). We measured K2-25's \lya~and X-ray flux and inferred the EUV flux in order to determine the planetary irradiation. Although all three planets experience EUV irradiation on the same order of magnitude, K2-25b experiences enhanced X-ray irradiation compared to the two other hot Neptunes. This is a result of K2-25's young age, and the slower decline of EUV with age compared to X-rays \citep{2020ApJ...895....5P}. The large XUV irradiation results in an estimated maximum mass loss rate ($10.6^{+15.2}_{-6.13}\eta \times 10^{10}$ g s$^{-1}$) that is five times larger than for Gl 436b \citep[{$2.20\eta \times10^{10}$ g/s};][]{2016A&A...591A.121B}. The larger mass loss rate for K2-25b means that if the same atmospheric escape process translates across all three planets, then we should be able to observe an escaping neutral hydrogen exosphere surrounding K2-25b.

Our analysis of two {\it HST}/STIS observations of K2-25 in \lya~yielded a non-detection of an extended neutral hydrogen envelope around K2-25b despite the large mass loss rate. This suggests that K2-25b is either not experiencing significant atmospheric escape, or other factors result in mass loss that is unobservable in our \lya~observations. The latter could be due to properties of the system itself. While adopting the $24.5$\Mearth~mass in place of the value from mass-radius relations results in a factor of three lower mass-loss rate, this reduction is not sufficient to imply a non-detection of an evaporating atmosphere, and we therefore explore the latter possibility.

\subsection{Astrophysical causes for a non-detection}

A possible explanation for the suppression of a neutral hydrogen exosphere is a denser, non-hydrogen dominated atmosphere. A higher mean molecular weight would decrease the atmospheric scale height, reducing the likelihood of hydrodynamic escape. This could be possible if K2-25b lost its primordial atmosphere when it was younger, which would be consistent with the potentially short $\sim100$ Myr timescale of photoevaporation and inconsistent with the longer Gyr timescale proposed by \cite{2021MNRAS.501L..28K}. \cite{2020ApJ...888L..21G} suggest a transition around $\rho_{\text{p}} \sim 2-3$ g cm$^{-3}$: planets with bulk densities below 2~g~cm$^{-3}$ have neutral hydrogen-dominated atmospheres that allow the detection of an extended atmosphere in \lya. Planets with bulk densities above 3~g~cm$^{-3}$ could have atmospheres dominated by heavier species, resulting in no detection of atmospheric absorption in \lya~regardless of mass loss rate. With the planetary mass estimate based on mass-radius relations ($7^{+10}_{-4}$\Mearth), K2-25b has a bulk density ($\rho_{\text{p}} = 0.94^{+1.35}_{-0.54}$~g~cm$^{-3}$) that places it within the regime of H-dominated atmospheres that should result in \lya~absorption if the atmosphere is escaping. However, the measurement from \cite{2020AJ....160..192S} results in a bulk density that agrees with either a H-dominated or water-dominated atmosphere ($\rho_{\text{p}} = 3.28\pm 0.8$~g~cm$^{-3}$). In the framework of \cite{2020ApJ...888L..21G}, K2-25b could potentially then fall into the regime where planets typically do not exhibit \lya~absorption even with high mass loss rates.

K2-25's youth could also impact the detectability of the planet's \lya~transit. The estimated lifetime before ionization of an escaping neutral hydrogen atom in K2-25b's exosphere is short ($2.86$ hours) compared to Gl 436b ($\sim$12 hours). The short lifetime could indicate that enough neutral atoms are ionized before they are able to travel far enough to form a large neutral hydrogen exosphere. The result may be a smaller exosphere more similar to that of GJ 3470. Our observational strategy was set prior to the detection of GJ 3470, and was optimized for the detection of a more extended exosphere. 

The exosphere could be interacting with high velocity stellar winds from its host \citep[e.g.,][]{2015ApJ...806...41C,2014A&A...562A.116K,2013A&A...557A.124B}. Via charge exchange, fast-traveling protons from the stellar wind may receive an electron from the neutral hydrogen in the transiting planet's atmosphere, suffering little deflection or change in kinetic energy from the collision. Therefore, if the stellar wind is high speed, the now-neutral stellar wind could result in \lya~absorption beyond the \lya~emission feature--where, for K2-25, there is little signal \citep[cf.][]{2016A&A...591A.121B}.

\cite{2021MNRAS.500.3382C} explored the effects of varying stellar wind strengths on the \lya~transits of close-in planets with 3D hydrodynamic simulations (see also \cite{2020MNRAS.498L..53C} and \cite{2020MNRAS.494.2417V}). They showed that stellar winds provide an external pressure that can confine the atmosphere of an exoplanet and decrease the exosphere's transit depth by a factor of about two. We make a rough estimate of the stellar mass loss rate for K2-25 using its X-ray flux ($3.17 \times 10^6$ erg s$^{-1}$ cm$^{-2}$) and the observed relationship between X-ray flux and the mass loss rates of main sequence stars \citep[c.f. Fig 3]{2005ApJ...628L.143W}. This gives $\sim0.8\dot{\text{M}}_{\odot}$. When making this estimate, we assume that the mass loss rate per unit surface area reaches an asymptote at $\sim9\dot{\text{M}}_{\odot}/\text{A}_{\odot}$ (units of solar mass loss rate per solar surface area) for X-ray fluxes greater than $10^6$ erg s$^{-1}$ cm$^{-2}$. The inferred mass loss rate for K2-25 is significantly lower than the wind strengths from \cite{2021MNRAS.500.3382C} and so we do not a priori expect that the wind would inhibit the detection of an escaping exosphere. Our stellar mass loss estimate is highly uncertain, however, as there are no observational constraints on K2-25's stellar wind strength and current $\dot{\text{M}_{\star}}\propto F_{X}$ relations do not extend to the high X-ray fluxes exhibited by young M dwarfs.

The stellar and planetary properties that concern the properties of evaporating atmospheres are far from clear, and K2-25b is currently the only young Neptune in the literature with an analysis of its \lya\ transit. Our non-detection of an extended atmosphere is inconsistent with the extended, cometary-tail seen around Gl 436b, but permits the presence of smaller exospheres more similar to that of GJ 3470b. We considered the scenario where we obtained a total of 3 {\it HST} visits similar in quality to our Visit 2. We can only detect an exosphere of $>0.5$R$_{\star}$ in the \lya\ blue-wing, which does not enable us to probe atmospheres as small as could be expected. K2-25b needs further modeling to learn about the expected size and shape of a potential exosphere, which could then guide future observations.

\acknowledgments
We sincerely thank Dr. Hans R. M\"uller and Dr. Brian Chaboyer for helpful discussions and thoughtful comments throughout the progress of this work. The authors appreciate the support, care, and community provided by the astronomy graduate students and post-docs within Dartmouth College's Department of Physics \& Astronomy. We would like to express our appreciation for Nova, Charlie, Edmund, and Nessie who have gifted us with smiles during many months of remote work.

This research is based on observations made with the NASA/ESA {\it Hubble Space Telescope} obtained from the Space Telescope Science Institute, which is operated by the Association of Universities for Research in Astronomy, Inc., under NASA contract NAS 5–26555. These observations are associated with HST-GO-14615. Support for program HST-GO-14615 was provided by NASA through a grant from the STScI.

This work is based on observations obtained with {\it XMM-Newton}, an ESA science mission with instruments and contributions directly funded by ESA Member States and NASA. These observations are associated with OBSID 0782061001.

This material is based upon work supported by the National Science Foundation under Grant No. 2008066.

This work benefited from the Exoplanet Summer Program in the Other Worlds Laboratory (OWL) at the University of California, Santa Cruz, a program funded by the Heising-Simons Foundation.

This project has received funding from the European Research Council (ERC) under the European Union’s Horizon 2020 research and innovation programme (project {\sc Four Aces}; grant agreement No 724427; project {\sc Spice Dune}; grant agreement No 947634). V.B. acknowledges support by the Swiss National Science Foundation (SNSF) in the frame of the National Centre for Competence in Research ``PlanetS''.

\vspace{5mm}
\facility{{\it HST}/STIS, {\it XMM-Newton}}

\software{
        {\tt astropy} \citep{astropy:2018},~
        {\tt batman} \citep{2015PASP..127.1161K},~
        {\tt calstis} (\url{https://github.com/spacetelescope/stistools}),~
        {\tt emcee} \citep{2013ApJ...763..149F},~
        {\tt lyapy} \citep{2016ApJ...824..101Y},~
        {\tt matplotlib} \citep{hunter2007matplotlib},~
        {\tt numpy} \citep{harris2020array},~
        {\tt scipy} \citep{2020SciPy}
        }

\newpage
\bibliography{references}

\begin{thebibliography}{}
\expandafter\ifx\csname natexlab\endcsname\relax\def\natexlab#1{#1}\fi
\providecommand{\url}[1]{\href{#1}{#1}}

\bibitem[{{Beaug{\'e}} \& {Nesvorn{\'y}}(2013)}]{2013ApJ...763...12B}
{Beaug{\'e}}, C., \& {Nesvorn{\'y}}, D. 2013, \apj, 763, 12

\bibitem[{{Ben-Jaffel} \& {Sona Hosseini}(2010)}]{2010ApJ...709.1284B}
{Ben-Jaffel}, L., \& {Sona Hosseini}, S. 2010, \apj, 709, 1284

\bibitem[{{Bourrier} {et~al.}(2015){Bourrier}, {Ehrenreich}, \& {Lecavelier des
  Etangs}}]{2015A&A...582A..65B}
{Bourrier}, V., {Ehrenreich}, D., \& {Lecavelier des Etangs}, A. 2015, \aap,
  582, A65

\bibitem[{{Bourrier} \& {Lecavelier des Etangs}(2013)}]{2013A&A...557A.124B}
{Bourrier}, V., \& {Lecavelier des Etangs}, A. 2013, \aap, 557, A124

\bibitem[{{Bourrier} {et~al.}(2016){Bourrier}, {Lecavelier des Etangs},
  {Ehrenreich}, {Tanaka}, \& {Vidotto}}]{2016A&A...591A.121B}
{Bourrier}, V., {Lecavelier des Etangs}, A., {Ehrenreich}, D., {Tanaka}, Y.~A.,
  \& {Vidotto}, A.~A. 2016, \aap, 591, A121

\bibitem[{{Bourrier} {et~al.}(2013){Bourrier}, {Lecavelier des Etangs},
  {Dupuy}, {Ehrenreich}, {Vidal-Madjar}, {H{\'e}brard}, {Ballester},
  {D{\'e}sert}, {Ferlet}, {Sing}, \& {Wheatley}}]{2013A&A...551A..63B}
{Bourrier}, V., {Lecavelier des Etangs}, A., {Dupuy}, H., {et~al.} 2013, \aap,
  551, A63

\bibitem[{{Bourrier} {et~al.}(2018{\natexlab{a}}){Bourrier}, {Lecavelier des
  Etangs}, {Ehrenreich}, {Sanz-Forcada}, {Allart}, {Ballester}, {Buchhave},
  {Cohen}, {Deming}, {Evans}, {Garc{\'\i}a Mu{\~n}oz}, {Henry}, {Kataria},
  {Lavvas}, {Lewis}, {L{\'o}pez-Morales}, {Marley}, {Sing}, \&
  {Wakeford}}]{2018A&A...620A.147B}
{Bourrier}, V., {Lecavelier des Etangs}, A., {Ehrenreich}, D., {et~al.}
  2018{\natexlab{a}}, \aap, 620, A147

\bibitem[{{Bourrier} {et~al.}(2018{\natexlab{b}}){Bourrier}, {Lovis}, {Beust},
  {Ehrenreich}, {Henry}, {Astudillo-Defru}, {Allart}, {Bonfils},
  {S{\'e}gransan}, {Delfosse}, {Cegla}, {Wyttenbach}, {Heng}, {Lavie}, \&
  {Pepe}}]{2018Natur.553..477B}
{Bourrier}, V., {Lovis}, C., {Beust}, H., {et~al.} 2018{\natexlab{b}}, \nat,
  553, 477

\bibitem[{{Bowers}(1997)}]{1997hstc.work...18B}
{Bowers}, C.~W. 1997, in The 1997 HST Calibration Workshop with a New
  Generation of Instruments, ed. S.~{Casertano}, R.~{Jedrzejewski}, T.~{Keyes},
  \& M.~{Stevens}, 18

\bibitem[{{Brown} {et~al.}(2001){Brown}, {Charbonneau}, {Gilliland}, {Noyes},
  \& {Burrows}}]{2001ApJ...552..699B}
{Brown}, T.~M., {Charbonneau}, D., {Gilliland}, R.~L., {Noyes}, R.~W., \&
  {Burrows}, A. 2001, \apj, 552, 699

\bibitem[{{Bzowski} {et~al.}(2013){Bzowski}, {Sok{\'o}{\l}}, {Tokumaru},
  {Fujiki}, {Qu{\'e}merais}, {Lallement}, {Ferron}, {Bochsler}, \&
  {McComas}}]{2013ccfu.book...67B}
{Bzowski}, M., {Sok{\'o}{\l}}, J.~M., {Tokumaru}, M., {et~al.} 2013, {Solar
  Parameters for Modeling the Interplanetary Background}, Vol.~13 (Springer
  Science+Business Media New York), 67

\bibitem[{{Carolan} {et~al.}(2020){Carolan}, {Vidotto}, {Plavchan}, {Villarreal
  D'Angelo}, \& {Hazra}}]{2020MNRAS.498L..53C}
{Carolan}, S., {Vidotto}, A.~A., {Plavchan}, P., {Villarreal D'Angelo}, C., \&
  {Hazra}, G. 2020, \mnras, 498, L53

\bibitem[{{Carolan} {et~al.}(2021){Carolan}, {Vidotto}, {Villarreal D'Angelo},
  \& {Hazra}}]{2021MNRAS.500.3382C}
{Carolan}, S., {Vidotto}, A.~A., {Villarreal D'Angelo}, C., \& {Hazra}, G.
  2021, \mnras, 500, 3382

\bibitem[{{Cohen} {et~al.}(2015){Cohen}, {Ma}, {Drake}, {Glocer}, {Garraffo},
  {Bell}, \& {Gombosi}}]{2015ApJ...806...41C}
{Cohen}, O., {Ma}, Y., {Drake}, J.~J., {et~al.} 2015, \apj, 806, 41

\bibitem[{{Davis} \& {Wheatley}(2009)}]{2009MNRAS.396.1012D}
{Davis}, T.~A., \& {Wheatley}, P.~J. 2009, \mnras, 396, 1012

\bibitem[{{Debrecht} {et~al.}(2020){Debrecht}, {Carroll-Nellenback}, {Frank},
  {Blackman}, {Fossati}, {McCann}, \& {Murray-Clay}}]{2020MNRAS.493.1292D}
{Debrecht}, A., {Carroll-Nellenback}, J., {Frank}, A., {et~al.} 2020, \mnras,
  493, 1292

\bibitem[{{dos Santos} {et~al.}(2019){dos Santos}, {Ehrenreich}, {Bourrier},
  {Lecavelier des Etangs}, {L{\'o}pez-Morales}, {Sing}, {Ballester},
  {Ben-Jaffel}, {Buchhave}, {Garc{\'\i}a Mu{\~n}oz}, {Henry}, {Kataria},
  {Lavie}, {Lavvas}, {Lewis}, {Mikal-Evans}, {Sanz-Forcada}, \&
  {Wakeford}}]{2019A&A...629A..47D}
{dos Santos}, L.~A., {Ehrenreich}, D., {Bourrier}, V., {et~al.} 2019, \aap,
  629, A47

\bibitem[{{Douglas} {et~al.}(2019){Douglas}, {Curtis}, {Ag{\"u}eros},
  {Cargile}, {Brewer}, {Meibom}, \& {Jansen}}]{2019ApJ...879..100D}
{Douglas}, S.~T., {Curtis}, J.~L., {Ag{\"u}eros}, M.~A., {et~al.} 2019, \apj,
  879, 100

\bibitem[{{Ehrenreich} {et~al.}(2015){Ehrenreich}, {Bourrier}, {Wheatley},
  {Lecavelier des Etangs}, {H{\'e}brard}, {Udry}, {Bonfils}, {Delfosse},
  {D{\'e}sert}, {Sing}, \& {Vidal-Madjar}}]{2015Natur.522..459E}
{Ehrenreich}, D., {Bourrier}, V., {Wheatley}, P.~J., {et~al.} 2015, \nat, 522,
  459

\bibitem[{{Erkaev} {et~al.}(2007){Erkaev}, {Kulikov}, {Lammer}, {Selsis},
  {Langmayr}, {Jaritz}, \& {Biernat}}]{2007A&A...472..329E}
{Erkaev}, N.~V., {Kulikov}, Y.~N., {Lammer}, H., {et~al.} 2007, \aap, 472, 329

\bibitem[{{Fontenla} {et~al.}(2014){Fontenla}, {Landi}, {Snow}, \&
  {Woods}}]{2014SoPh..289..515F}
{Fontenla}, J.~M., {Landi}, E., {Snow}, M., \& {Woods}, T. 2014, \solphys, 289,
  515

\bibitem[{{Foreman-Mackey} {et~al.}(2013){Foreman-Mackey}, {Hogg}, {Lang}, \&
  {Goodman}}]{2013PASP..125..306F}
{Foreman-Mackey}, D., {Hogg}, D.~W., {Lang}, D., \& {Goodman}, J. 2013,
  Publications of the Astronomical Society of the Pacific, 125, 306

\bibitem[{{France} {et~al.}(2013){France}, {Froning}, {Linsky}, {Roberge},
  {Stocke}, {Tian}, {Bushinsky}, {D{\'e}sert}, {Mauas}, {Vieytes}, \&
  {Walkowicz}}]{2013ApJ...763..149F}
{France}, K., {Froning}, C.~S., {Linsky}, J.~L., {et~al.} 2013, \apj, 763, 149

\bibitem[{{Fulton} \& {Petigura}(2018)}]{2018AJ....156..264F}
{Fulton}, B.~J., \& {Petigura}, E.~A. 2018, \aj, 156, 264

\bibitem[{{Fulton} {et~al.}(2017){Fulton}, {Petigura}, {Howard}, {Isaacson},
  {Marcy}, {Cargile}, {Hebb}, {Weiss}, {Johnson}, {Morton}, {Sinukoff},
  {Crossfield}, \& {Hirsch}}]{2017AJ....154..109F}
{Fulton}, B.~J., {Petigura}, E.~A., {Howard}, A.~W., {et~al.} 2017, \aj, 154,
  109

\bibitem[{{Gaidos} {et~al.}(2020){Gaidos}, {Hirano}, {Wilson}, {France},
  {Rockcliffe}, {Newton}, {Feiden}, {Krishnamurthy}, {Harakawa}, {Hodapp},
  {Ishizuka}, {Jacobson}, {Konishi}, {Kotani}, {Kudo}, {Kurokawa}, {Kuzuhara},
  {Nishikawa}, {Omiya}, {Serizawa}, {Tamura}, {Ueda}, \&
  {Vievard}}]{2020MNRAS.498L.119G}
{Gaidos}, E., {Hirano}, T., {Wilson}, D.~J., {et~al.} 2020, \mnras, 498, L119

\bibitem[{{Garc{\'\i}a Mu{\~n}oz} {et~al.}(2020){Garc{\'\i}a Mu{\~n}oz},
  {Youngblood}, {Fossati}, {Gandolfi}, {Cabrera}, \&
  {Rauer}}]{2020ApJ...888L..21G}
{Garc{\'\i}a Mu{\~n}oz}, A., {Youngblood}, A., {Fossati}, L., {et~al.} 2020,
  \apjl, 888, L21

\bibitem[{{Gehrels}(1986)}]{1986ApJ...303..336G}
{Gehrels}, N. 1986, \apj, 303, 336

\bibitem[{{Ginzburg} {et~al.}(2018){Ginzburg}, {Schlichting}, \&
  {Sari}}]{2018MNRAS.476..759G}
{Ginzburg}, S., {Schlichting}, H.~E., \& {Sari}, R. 2018, \mnras, 476, 759

\bibitem[{{Gossage} {et~al.}(2018){Gossage}, {Conroy}, {Dotter}, {Choi},
  {Rosenfield}, {Cargile}, \& {Dolphin}}]{2018ApJ...863...67G}
{Gossage}, S., {Conroy}, C., {Dotter}, A., {et~al.} 2018, \apj, 863, 67

\bibitem[{Harris {et~al.}(2020)Harris, Millman, van~der Walt, Gommers,
  Virtanen, Cournapeau, Wieser, Taylor, Berg, Smith, Kern, Picus, Hoyer, van
  Kerkwijk, Brett, Haldane, del R{'{\i}}o, Wiebe, Peterson,
  G{'{e}}rard-Marchant, Sheppard, Reddy, Weckesser, Abbasi, Gohlke, \&
  Oliphant}]{harris2020array}
Harris, C.~R., Millman, K.~J., van~der Walt, S.~J., {et~al.} 2020, Nature, 585,
  357.
\newblock \url{https://doi.org/10.1038/s41586-020-2649-2}

\bibitem[{{Harris}(1948)}]{1948ApJ...108..112H}
{Harris}, III, D.~L. 1948, \apj, 108, 112

\bibitem[{{H{\'e}brard} \& {Moos}(2003)}]{2003ApJ...599..297H}
{H{\'e}brard}, G., \& {Moos}, H.~W. 2003, \apj, 599, 297

\bibitem[{{Holmstr{\"o}m} {et~al.}(2008){Holmstr{\"o}m}, {Ekenb{\"a}ck},
  {Selsis}, {Penz}, {Lammer}, \& {Wurz}}]{2008Natur.451..970H}
{Holmstr{\"o}m}, M., {Ekenb{\"a}ck}, A., {Selsis}, F., {et~al.} 2008, \nat,
  451, 970

\bibitem[{{Huitson} {et~al.}(2012){Huitson}, {Sing}, {Vidal-Madjar},
  {Ballester}, {Lecavelier des Etangs}, {D{\'e}sert}, \&
  {Pont}}]{2012MNRAS.422.2477H}
{Huitson}, C.~M., {Sing}, D.~K., {Vidal-Madjar}, A., {et~al.} 2012, \mnras,
  422, 2477

\bibitem[{Hunter(2007)}]{hunter2007matplotlib}
Hunter, J.~D. 2007, Computing in science \& engineering, 9, 90

\bibitem[{{Jin} {et~al.}(2014){Jin}, {Mordasini}, {Parmentier}, {van Boekel},
  {Henning}, \& {Ji}}]{2014ApJ...795...65J}
{Jin}, S., {Mordasini}, C., {Parmentier}, V., {et~al.} 2014, \apj, 795, 65

\bibitem[{{Kain} {et~al.}(2020){Kain}, {Newton}, {Dittmann}, {Irwin}, {Mann},
  {Thao}, {Charbonneau}, \& {Winters}}]{2020AJ....159...83K}
{Kain}, I.~J., {Newton}, E.~R., {Dittmann}, J.~A., {et~al.} 2020, \aj, 159, 83

\bibitem[{{Kanodia} {et~al.}(2019){Kanodia}, {Wolfgang}, {Stefansson}, {Ning},
  \& {Mahadevan}}]{2019ascl.soft12020K}
{Kanodia}, S., {Wolfgang}, A., {Stefansson}, G.~K., {Ning}, B., \& {Mahadevan},
  S. 2019, {MRExo: Non-parametric mass-radius relationship for exoplanets},
  Astrophysics Source Code Library, ascl:1912.020

\bibitem[{{King} \& {Wheatley}(2021)}]{2021MNRAS.501L..28K}
{King}, G.~W., \& {Wheatley}, P.~J. 2021, \mnras, 501, L28

\bibitem[{{Kislyakova} {et~al.}(2014){Kislyakova}, {Johnstone}, {Odert},
  {Erkaev}, {Lammer}, {L{\"u}ftinger}, {Holmstr{\"o}m}, {Khodachenko}, \&
  {G{\"u}del}}]{2014A&A...562A.116K}
{Kislyakova}, K.~G., {Johnstone}, C.~P., {Odert}, P., {et~al.} 2014, \aap, 562,
  A116

\bibitem[{{Kosiarek} {et~al.}(2019){Kosiarek}, {Crossfield},
  {Hardegree-Ullman}, {Livingston}, {Benneke}, {Henry}, {Howard}, {Berardo},
  {Blunt}, {Fulton}, {Hirsch}, {Howard}, {Isaacson}, {Petigura}, {Sinukoff},
  {Weiss}, {Bonfils}, {Dressing}, {Knutson}, {Schlieder}, {Werner}, {Gorjian},
  {Krick}, {Morales}, {Astudillo-Defru}, {Almenara}, {Delfosse}, {Forveille},
  {Lovis}, {Mayor}, {Murgas}, {Pepe}, {Santos}, {Udry}, {Corbett}, {Fors},
  {Law}, {Ratzloff}, \& {del Ser}}]{2019AJ....157...97K}
{Kosiarek}, M.~R., {Crossfield}, I. J.~M., {Hardegree-Ullman}, K.~K., {et~al.}
  2019, \aj, 157, 97

\bibitem[{{Kreidberg}(2015)}]{2015PASP..127.1161K}
{Kreidberg}, L. 2015, \pasp, 127, 1161

\bibitem[{{Kulow} {et~al.}(2014){Kulow}, {France}, {Linsky}, \&
  {Loyd}}]{2014ApJ...786..132K}
{Kulow}, J.~R., {France}, K., {Linsky}, J., \& {Loyd}, R.~O.~P. 2014, \apj,
  786, 132

\bibitem[{{Kurokawa} \& {Nakamoto}(2014)}]{2014ApJ...783...54K}
{Kurokawa}, H., \& {Nakamoto}, T. 2014, \apj, 783, 54

\bibitem[{{Lammer} {et~al.}(2003){Lammer}, {Selsis}, {Ribas}, {Guinan},
  {Bauer}, \& {Weiss}}]{2003ApJ...598L.121L}
{Lammer}, H., {Selsis}, F., {Ribas}, I., {et~al.} 2003, \apjl, 598, L121

\bibitem[{{Lavie} {et~al.}(2017){Lavie}, {Ehrenreich}, {Bourrier}, {Lecavelier
  des Etangs}, {Vidal-Madjar}, {Delfosse}, {Gracia Berna}, {Heng}, {Thomas},
  {Udry}, \& {Wheatley}}]{2017A&A...605L...7L}
{Lavie}, B., {Ehrenreich}, D., {Bourrier}, V., {et~al.} 2017, \aap, 605, L7

\bibitem[{{Lecavelier Des Etangs}(2007)}]{2007A&A...461.1185L}
{Lecavelier Des Etangs}, A. 2007, \aap, 461, 1185

\bibitem[{{Lee} \& {Connors}(2021)}]{2021ApJ...908...32L}
{Lee}, E.~J., \& {Connors}, N.~J. 2021, \apj, 908, 32

\bibitem[{{Linsky} {et~al.}(2014){Linsky}, {Fontenla}, \&
  {France}}]{2014ApJ...780...61L}
{Linsky}, J.~L., {Fontenla}, J., \& {France}, K. 2014, \apj, 780, 61

\bibitem[{{Linsky} {et~al.}(2006){Linsky}, {Draine}, {Moos}, {Jenkins}, {Wood},
  {Oliveira}, {Blair}, {Friedman}, {Gry}, {Knauth}, {Kruk}, {Lacour}, {Lehner},
  {Redfield}, {Shull}, {Sonneborn}, \& {Williger}}]{2006ApJ...647.1106L}
{Linsky}, J.~L., {Draine}, B.~T., {Moos}, H.~W., {et~al.} 2006, \apj, 647, 1106

\bibitem[{{Lopez} \& {Fortney}(2013)}]{2013ApJ...776....2L}
{Lopez}, E.~D., \& {Fortney}, J.~J. 2013, \apj, 776, 2

\bibitem[{{Loyd} {et~al.}(2018){Loyd}, {France}, {Youngblood}, {Schneider},
  {Brown}, {Hu}, {Segura}, {Linsky}, {Redfield}, {Tian}, {Rugheimer}, {Miguel},
  \& {Froning}}]{2018ApJ...867...71L}
{Loyd}, R.~O.~P., {France}, K., {Youngblood}, A., {et~al.} 2018, \apj, 867, 71

\bibitem[{{Lundkvist} {et~al.}(2016){Lundkvist}, {Kjeldsen}, {Albrecht},
  {Davies}, {Basu}, {Huber}, {Justesen}, {Karoff}, {Silva Aguirre}, {van
  Eylen}, {Vang}, {Arentoft}, {Barclay}, {Bedding}, {Campante}, {Chaplin},
  {Christensen-Dalsgaard}, {Elsworth}, {Gilliland}, {Handberg}, {Hekker},
  {Kawaler}, {Lund}, {Metcalfe}, {Miglio}, {Rowe}, {Stello}, {Tingley}, \&
  {White}}]{2016NatCo...711201L}
{Lundkvist}, M.~S., {Kjeldsen}, H., {Albrecht}, S., {et~al.} 2016, Nature
  Communications, 7, 11201

\bibitem[{{Mann} {et~al.}(2016){Mann}, {Gaidos}, {Mace}, {Johnson}, {Bowler},
  {LaCourse}, {Jacobs}, {Vanderburg}, {Kraus}, {Kaplan}, \&
  {Jaffe}}]{2016ApJ...818...46M}
{Mann}, A.~W., {Gaidos}, E., {Mace}, G.~N., {et~al.} 2016, \apj, 818, 46

\bibitem[{{Mazeh} {et~al.}(2016){Mazeh}, {Holczer}, \&
  {Faigler}}]{2016A&A...589A..75M}
{Mazeh}, T., {Holczer}, T., \& {Faigler}, S. 2016, \aap, 589, A75

\bibitem[{{McCann} {et~al.}(2019){McCann}, {Murray-Clay}, {Kratter}, \&
  {Krumholz}}]{2019ApJ...873...89M}
{McCann}, J., {Murray-Clay}, R.~A., {Kratter}, K., \& {Krumholz}, M.~R. 2019,
  \apj, 873, 89

\bibitem[{{Murray-Clay} {et~al.}(2009){Murray-Clay}, {Chiang}, \&
  {Murray}}]{2009ApJ...693...23M}
{Murray-Clay}, R.~A., {Chiang}, E.~I., \& {Murray}, N. 2009, \apj, 693, 23

\bibitem[{{Ning} {et~al.}(2018){Ning}, {Wolfgang}, \&
  {Ghosh}}]{2018ApJ...869....5N}
{Ning}, B., {Wolfgang}, A., \& {Ghosh}, S. 2018, \apj, 869, 5

\bibitem[{{Owen}(2019)}]{2019AREPS..47...67O}
{Owen}, J.~E. 2019, Annual Review of Earth and Planetary Sciences, 47, 67

\bibitem[{{Owen} \& {Jackson}(2012)}]{2012MNRAS.425.2931O}
{Owen}, J.~E., \& {Jackson}, A.~P. 2012, \mnras, 425, 2931

\bibitem[{{Owen} \& {Lai}(2018)}]{2018MNRAS.479.5012O}
{Owen}, J.~E., \& {Lai}, D. 2018, \mnras, 479, 5012

\bibitem[{{Owen} \& {Wu}(2013)}]{2013ApJ...775..105O}
{Owen}, J.~E., \& {Wu}, Y. 2013, \apj, 775, 105

\bibitem[{{Owen} \& {Wu}(2017)}]{2017ApJ...847...29O}
---. 2017, \apj, 847, 29

\bibitem[{{Peacock} {et~al.}(2019){Peacock}, {Barman}, {Shkolnik},
  {Hauschildt}, {Baron}, \& {Fuhrmeister}}]{2019ApJ...886...77P}
{Peacock}, S., {Barman}, T., {Shkolnik}, E.~L., {et~al.} 2019, \apj, 886, 77

\bibitem[{{Peacock} {et~al.}(2020){Peacock}, {Barman}, {Shkolnik}, {Loyd},
  {Schneider}, {Pagano}, \& {Meadows}}]{2020ApJ...895....5P}
---. 2020, \apj, 895, 5

\bibitem[{{Price-Whelan} {et~al.}(2018){Price-Whelan}, {Sip{\H{o}}cz},
  {G{\"u}nther}, {Lim}, {Crawford}, {Conseil}, {Shupe}, {Craig}, {Dencheva},
  {Ginsburg}, {VanderPlas}, {Bradley}, {P{\'e}rez-Su{\'a}rez}, {de Val-Borro},
  {Paper Contributors}, {Aldcroft}, {Cruz}, {Robitaille}, {Tollerud},
  {Coordination Committee}, {Ardelean}, {Babej}, {Bach}, {Bachetti}, {Bakanov},
  {Bamford}, {Barentsen}, {Barmby}, {Baumbach}, {Berry}, {Biscani}, {Boquien},
  {Bostroem}, {Bouma}, {Brammer}, {Bray}, {Breytenbach}, {Buddelmeijer},
  {Burke}, {Calderone}, {Cano Rodr{\'\i}guez}, {Cara}, {Cardoso}, {Cheedella},
  {Copin}, {Corrales}, {Crichton}, {D{\textquoteright}Avella}, {Deil},
  {Depagne}, {Dietrich}, {Donath}, {Droettboom}, {Earl}, {Erben}, {Fabbro},
  {Ferreira}, {Finethy}, {Fox}, {Garrison}, {Gibbons}, {Goldstein}, {Gommers},
  {Greco}, {Greenfield}, {Groener}, {Grollier}, {Hagen}, {Hirst}, {Homeier},
  {Horton}, {Hosseinzadeh}, {Hu}, {Hunkeler}, {Ivezi{\'c}}, {Jain}, {Jenness},
  {Kanarek}, {Kendrew}, {Kern}, {Kerzendorf}, {Khvalko}, {King}, {Kirkby},
  {Kulkarni}, {Kumar}, {Lee}, {Lenz}, {Littlefair}, {Ma}, {Macleod},
  {Mastropietro}, {McCully}, {Montagnac}, {Morris}, {Mueller}, {Mumford},
  {Muna}, {Murphy}, {Nelson}, {Nguyen}, {Ninan}, {N{\"o}the}, {Ogaz}, {Oh},
  {Parejko}, {Parley}, {Pascual}, {Patil}, {Patil}, {Plunkett}, {Prochaska},
  {Rastogi}, {Reddy Janga}, {Sabater}, {Sakurikar}, {Seifert}, {Sherbert},
  {Sherwood-Taylor}, {Shih}, {Sick}, {Silbiger}, {Singanamalla}, {Singer},
  {Sladen}, {Sooley}, {Sornarajah}, {Streicher}, {Teuben}, {Thomas},
  {Tremblay}, {Turner}, {Terr{\'o}n}, {van Kerkwijk}, {de la Vega}, {Watkins},
  {Weaver}, {Whitmore}, {Woillez}, {Zabalza}, \& {Contributors}}]{astropy:2018}
{Price-Whelan}, A.~M., {Sip{\H{o}}cz}, B.~M., {G{\"u}nther}, H.~M., {et~al.}
  2018, \aj, 156, 123

\bibitem[{{Redfield} \& {Linsky}(2000)}]{2000ApJ...534..825R}
{Redfield}, S., \& {Linsky}, J.~L. 2000, \apj, 534, 825

\bibitem[{{Redfield} \& {Linsky}(2008)}]{2008ApJ...673..283R}
---. 2008, \apj, 673, 283

\bibitem[{{Shaikhislamov} {et~al.}(2014){Shaikhislamov}, {Khodachenko},
  {Sasunov}, {Lammer}, {Kislyakova}, \& {Erkaev}}]{2014ApJ...795..132S}
{Shaikhislamov}, I.~F., {Khodachenko}, M.~L., {Sasunov}, Y.~L., {et~al.} 2014,
  \apj, 795, 132

\bibitem[{{Sing} {et~al.}(2008){Sing}, {Vidal-Madjar}, {D{\'e}sert},
  {Lecavelier des Etangs}, \& {Ballester}}]{2008ApJ...686..658S}
{Sing}, D.~K., {Vidal-Madjar}, A., {D{\'e}sert}, J.~M., {Lecavelier des
  Etangs}, A., \& {Ballester}, G. 2008, \apj, 686, 658

\bibitem[{{Stassun} {et~al.}(2019){Stassun}, {Oelkers}, {Paegert}, {Torres},
  {Pepper}, {De Lee}, {Collins}, {Latham}, {Muirhead}, {Chittidi},
  {Rojas-Ayala}, {Fleming}, {Rose}, {Tenenbaum}, {Ting}, {Kane}, {Barclay},
  {Bean}, {Brassuer}, {Charbonneau}, {Ge}, {Lissauer}, {Mann}, {McLean},
  {Mullally}, {Narita}, {Plavchan}, {Ricker}, {Sasselov}, {Seager}, {Sharma},
  {Shiao}, {Sozzetti}, {Stello}, {Vanderspek}, {Wallace}, \&
  {Winn}}]{2019AJ....158..138S}
{Stassun}, K.~G., {Oelkers}, R.~J., {Paegert}, M., {et~al.} 2019, \aj, 158, 138

\bibitem[{{Stefansson} {et~al.}(2020){Stefansson}, {Mahadevan}, {Maney},
  {Ninan}, {Robertson}, {Rajagopal}, {Haase}, {Allen}, {Ford}, {Winn},
  {Wolfgang}, {Dawson}, {Wisniewski}, {Bender}, {Ca{\~n}as}, {Cochran},
  {Diddams}, {Fredrick}, {Halverson}, {Hearty}, {Hebb}, {Kanodia}, {Levi},
  {Metcalf}, {Monson}, {Ramsey}, {Roy}, {Schwab}, {Terrien}, \&
  {Wright}}]{2020AJ....160..192S}
{Stefansson}, G., {Mahadevan}, S., {Maney}, M., {et~al.} 2020, \aj, 160, 192

\bibitem[{{Szab{\'o}} \& {Kiss}(2011)}]{2011ApJ...727L..44S}
{Szab{\'o}}, G.~M., \& {Kiss}, L.~L. 2011, \apjl, 727, L44

\bibitem[{{Thao} {et~al.}(2020){Thao}, {Mann}, {Johnson}, {Newton}, {Guo},
  {Kain}, {Rizzuto}, {Charbonneau}, {Dalba}, {Gaidos}, {Irwin}, \&
  {Kraus}}]{2020AJ....159...32T}
{Thao}, P.~C., {Mann}, A.~W., {Johnson}, M.~C., {et~al.} 2020, \aj, 159, 32

\bibitem[{{Tremblin} \& {Chiang}(2013)}]{2013MNRAS.428.2565T}
{Tremblin}, P., \& {Chiang}, E. 2013, \mnras, 428, 2565

\bibitem[{{Verner} {et~al.}(1996){Verner}, {Ferland}, {Korista}, \&
  {Yakovlev}}]{1996ApJ...465..487V}
{Verner}, D.~A., {Ferland}, G.~J., {Korista}, K.~T., \& {Yakovlev}, D.~G. 1996,
  \apj, 465, 487

\bibitem[{{Veyette} \& {Muirhead}(2018)}]{2018ApJ...863..166V}
{Veyette}, M.~J., \& {Muirhead}, P.~S. 2018, \apj, 863, 166

\bibitem[{{Vidotto} \& {Cleary}(2020)}]{2020MNRAS.494.2417V}
{Vidotto}, A.~A., \& {Cleary}, A. 2020, \mnras, 494, 2417

\bibitem[{{Villarreal D'Angelo} {et~al.}(2018){Villarreal D'Angelo},
  {Esquivel}, {Schneiter}, \& {Sgr{\'o}}}]{2018MNRAS.479.3115V}
{Villarreal D'Angelo}, C., {Esquivel}, A., {Schneiter}, M., \& {Sgr{\'o}},
  M.~A. 2018, \mnras, 479, 3115

\bibitem[{Virtanen {et~al.}(2020)Virtanen, Gommers, Oliphant, Haberland, Reddy,
  Cournapeau, Burovski, Peterson, {Weckesser}, {Bright}, {van der Walt},
  {Brett}, {Wilson}, {Jarrod Millman}, {Mayorov}, {Nelson}, {Jones}, {Kern},
  {Larson}, {Carey}, {Polat}, {Feng}, {Moore}, {Vand erPlas}, {Laxalde},
  {Perktold}, {Cimrman}, {Henriksen}, {Quintero}, {Harris}, {Archibald},
  {Ribeiro}, {Pedregosa}, {van Mulbregt}, \& {Contributors}}]{2020SciPy}
Virtanen, P., Gommers, R., Oliphant, T.~E., {et~al.} 2020, Nature Methods

\bibitem[{{Watson} {et~al.}(1981){Watson}, {Donahue}, \&
  {Walker}}]{1981Icar...48..150W}
{Watson}, A.~J., {Donahue}, T.~M., \& {Walker}, J.~C.~G. 1981, \icarus, 48, 150

\bibitem[{{Wood} {et~al.}(2004){Wood}, {Linsky}, {H{\'e}brard}, {Williger},
  {Moos}, \& {Blair}}]{2004ApJ...609..838W}
{Wood}, B.~E., {Linsky}, J.~L., {H{\'e}brard}, G., {et~al.} 2004, \apj, 609,
  838

\bibitem[{{Wood} {et~al.}(2005){Wood}, {M{\"u}ller}, {Zank}, {Linsky}, \&
  {Redfield}}]{2005ApJ...628L.143W}
{Wood}, B.~E., {M{\"u}ller}, H.~R., {Zank}, G.~P., {Linsky}, J.~L., \&
  {Redfield}, S. 2005, \apjl, 628, L143

\bibitem[{{Wright} {et~al.}(2018){Wright}, {Newton}, {Williams}, {Drake}, \&
  {Yadav}}]{2018MNRAS.479.2351W}
{Wright}, N.~J., {Newton}, E.~R., {Williams}, P. K.~G., {Drake}, J.~J., \&
  {Yadav}, R.~K. 2018, \mnras, 479, 2351

\bibitem[{{Yelle}(2004)}]{2004Icar..170..167Y}
{Yelle}, R.~V. 2004, \icarus, 170, 167

\bibitem[{{Youngblood} {et~al.}(2016){Youngblood}, {France}, {Loyd}, {Linsky},
  {Redfield}, {Schneider}, {Wood}, {Brown}, {Froning}, {Miguel}, {Rugheimer},
  \& {Walkowicz}}]{2016ApJ...824..101Y}
{Youngblood}, A., {France}, K., {Loyd}, R.~O.~P., {et~al.} 2016, \apj, 824, 101

\end{thebibliography}

\end{document}